\documentstyle[12pt,color,epsfig]{article}
\include{colordvi}
\begin{document}
\newcommand{\p}{\partial}
\newcommand{\ov}{\overline}
\newcommand{\da}{^{\dagger}}
\newcommand{\w}{\wedge}
\newcommand{\al}{\alpha}
\newcommand{\bb}{\beta}
\newcommand{\ga}{\gamma}
\newcommand{\te}{\theta}
\newcommand{\de}{\delta}
\newcommand{\et}{\eta}
\newcommand{\ze}{\zeta}
\newcommand{\s}{\sigma}
\newcommand{\e}{\epsilon}
\newcommand{\om}{\omega}
\newcommand{\la}{\lambda}
\newcommand{\La}{\Lambda}
\newcommand{\ad}{\dot{\alpha}}
\newcommand{\bd}{\dot{\beta}}
\newcommand{\gd}{\dot{\gamma}}
\newcommand{\dd}{\dot{\delta}}
\newcommand{\ed}{\dot{\eta}}
\newcommand{\zd}{\dot{\zeta}}
\newcommand{\md}{\dot{\mu}}
\newcommand{\nd}{\dot{\nu}}
\newcommand{\tb}{\overline{\theta}}
\newcommand{\D}{\overline{D}}
\newcommand{\Dap}{\overline{D}_{\ad +}}
\newcommand{\Dbp}{\overline{D}_{\bd +}}
\newcommand{\Dgp}{\overline{D}_{\gd +}}
\newcommand{\Ddp}{\overline{D}_{\dd +}}
\newcommand{\Dep}{\overline{D}_{\ed +}}
\newcommand{\Dzp}{\overline{D}_{\zd +}}
\newcommand{\Dam}{\overline{D}_{\ad -}}
\newcommand{\Dbm}{\overline{D}_{\bd -}}
\newcommand{\Dgm}{\overline{D}_{\gd -}}
\newcommand{\Ddm}{\overline{D}_{\dd -}}
\newcommand{\Dem}{\overline{D}_{\ed -}}
\newcommand{\Dzm}{\overline{D}_{\zd -}}

\newcommand{\ti}{\widetilde}

\newcommand{\2}{\frac{1}{2}}
\newcommand{\3}{\frac{1}{3}}
\newcommand{\4}{\frac{1}{4}}
\newcommand{\8}{\frac{1}{8}}
\newcommand{\6}{\frac{1}{16}}
\newcommand{\ra}{\rightarrow}
\newcommand{\Ra}{\Rightarrow}
\newcommand{\im}{\Longleftrightarrow}
\newcommand{\Dp}{D_{--}}
\newcommand{\Dm}{D_{++}}
\newcommand{\x}{\star}

\thispagestyle{empty}
{\bf 7th May, 2001} \hspace{\fill}
{\bf YITP-SB-01-19}

\vspace{1cm}
\begin{center}{\Huge{\bf N=2 HARMONIC\\

\vspace{3mm}
SUPERFORMS, MULTIPLETS\\

\vspace{5mm}
AND ACTIONS}}\\
\vspace{1cm}
{\large{\bf T. Biswas\footnote{tirtho@insti.physics.sunysb.edu} and W. Siegel\footnote{siegel@insti.physics.sunysb.edu}}}\\
\vspace{5mm}
{\small C.N. Yang Institute of Theoretical Physics,\\
Department of Physics and Astronomy,\\
State University of New York at Stony Brook,\\
Stony Brook, New York 11794-3840}
\\
\end{center}

\vspace{2cm}

\begin{abstract}
 In this paper we develop the formalism to study superforms in N=2 harmonic superspace. We perform a thorough (if not complete) analysis of the superforms starting from 0-form and moving all the way up to 6-form. Like the N=1 case we find that the lower superforms (0,1,2,3) describe the various important N=2 supermultiplets. Also, the forms form chains, the field strength of 0-form being related to the gauge form of the 1-form, and so on. However, an important difference with the N=1 case is that there is now more than one chain. 

Our main aim was to study the higher-forms (4,5,6) to obtain the various N=2 action formulas via the ectoplasmic approach. Indeed, we  reproduce the three known action formulas involving a volume integral, a contour integral and no integral over the harmonic subspace from the 6, 5 and 4-form analysis respectively. The next aim  is to generalize the analysis to curved space-time (supergravity).  
\end{abstract}

\newpage
\setcounter{page}{1}
\begin{center} {\bf {\large  INTRODUCTION} }
\end{center}

It is well known that not all  N=2 supermultiplets can conveniently be described 
off-shell using ordinary N=2 superspace. The scalar hypermultiplet is  a  case in point. Indeed, there is  a  no-go theorem  \cite{nogo} which states that any formulation of the scalar hypermultiplet involving  a  finite number of auxiliary fields will be 
inadequate to construct its off-shell representation. We can  bypass this theorem by enlarging the N=2 ordinary superspace, by adding an extra bosonic manifold, so that fields on this enlarged space can provide us with an infinite number of ``auxiliary fields'' (which depend only on  ordinary N=2 superspace co-ordinates) via harmonic (mode) expansions on the extra compact manifold. The most ``natural'' and successful among such ventures is based on harmonic superspace  \cite{harm} which consists of  a  sphere, or  technically speaking the coset SU(2)/U(1), in addition to the usual superspace co-ordinates. Indeed, in this harmonic superspace one finds adequate off-shell description of not only the vector (Yang-Mills) and the tensor multiplets but also the scalar multiplets. 

Considerable  work has now been done in this formalism, for example in understanding various dualities, harmonic supergraphs, N=2 supergravity theories, etc.  A  major obstacle however,  in formulating supergravity theories,  is writing down the action in terms of the supersymmetric (gauge invariant) objects (``field strengths'') because they  couple non trivially  to the vielbein  field. An elegant solution to the problem  can be achieved by using the so called ``ectoplasmic integration formula''  \cite{ecto} which was successfully applied to the N=1 case. This paper (hopefully) is the first step towards obtaining the N=2 supergravity action formula using ectoplasmic/superform techniques. Basically, the ectoplasmic approach  (see section 4 for details) relies on the fact that if we have a p-form, say, $A_p$,  which is closed ($dA_p=0$) but not exact (i.e. $A_p\neq d\La_{p-1}$), then the integral of the p-form over a p-dimensional subspace will be independent of all the co-ordinates. If we find such forms, then we can obtain actions (which also need to be independent of all the co-ordinates) in terms of the ``prepotential''(the superfield in terms of which we can express $A_p$) via the superform integral!  The superform integral will automatically contain the nontrivial couplings of the components of $A_p$ with the vielbein. In this paper we generate the already known action formulas for rigid harmonic N=2 superspace. There are 3 integration formulas known, one for chiral, one for constrained analytic (in a different language  \cite{cp11} though) and one for unconstrained analytic superfields, involving integration over zero, one or two harmonic (spherical) co-ordinates respectively. The 4,5 and 6 harmonic superforms generate these formulas as we will show. Moreover, it seems that there is a new ectoplasmic action formula for integrating unconstrained analytic superfields just over four dimensional (and not the conventional six) space-time  which perhaps suggests that there are many others, too. Here is a qualitative summary of the ectoplasmic action formulas that we obtained:
$$
\begin{array}{llll}
${\bf level}$ & ${\bf prepotential}$ & ${\bf nature}$ & ${\bf action}$\\
$4-form$& V&  $chiral$ & S= \int d^4x (\D_+^2\D_-^2V+c.c.)\\
$5-form$& V^{++}& $constrained$& S=\int d^4x\ \om^{++}  \ \D_+^2D^{-2}V^{++}\\
 & & $analytic$& \ =\int d^4x\oint\frac{dt}{t}D^4V(t)\\
$6-form$& V^{4+}&$unconstrained$& S=\int d^4x \ d^2y  D^{-2}\D_+^2 V^{4+}\\
 & & $analytic$& 
\end{array}
$$   
where $D,\bar{D}$ are the spinor covariant derivatives, $\om^{++}$ is a form in one of the spherical directions, and $(t,\bar{t})$ is a special co-ordinate system spanning the sphere, while $d^2y$ denotes surface integration over the sphere.

The lower-forms usually describe most of the important supermultiplets that exist in the theory.  For example, in N=1 the 0,1,2 and 3 forms describe the chiral scalar, Yang-Mills vector, chiral tensor and auxiliary (linear) multiplets respectively  \cite{n=1,siegel}. Moreover, the different forms are linked together in a chain-like fashion where the field strength of one resembles the gauge form of the next. We realized
 that a comprehensive treatment of the lower forms  in a systematic manner for harmonic superspace does not exist in the literature,  although people have studied several of them separately or in a different language. (For example  Yang-Mill's vector multiplet (1-form) has been studied in detail in the harmonic superspace language  \cite{harm}, while a formulation of the 2-form (tensor multiplet)  exists in ordinary superspace\footnote{Tensor multiplets have indeed been studied extensively using harmonic superfields, but their origin as a constrained (irreducible) 2-form in the harmonic superspace has not been discussed.}  \cite{ten}). As a warm-up and a review exercise, and because the lower forms help us to understand the higher forms (and vice-versa), we try in this paper to present  a  clear and structured analysis of all the superforms in harmonic superspace. In brief, here are the lower-form statistics:

\vspace{5mm}
\begin{tabular}{||l|ll|l|l||}\hline
{\bf level}& {\bf prepot.}& {\bf nature}& {\bf field strength}& {\bf describes}\\ \hline
0-form& $V$& unconstrained& analytic& scalar\\
 & & analytic & & multiplet\\
1-form& $V^{++}$&unconstrained& chiral& vector\\
 & & analytic & & multiplet\\
2-form& $V$& constrained& analytic & tensor\\
 & & chiral& & multiplet\\
3-forms& $V^{++}$& unconstrained& (a) analytic& 3-form\\
(2) & & analytic& (b) chiral& \\ \hline 
\end{tabular}

\vspace{5mm}

In section 1 we introduce harmonic superspace and develop the necessary tools to work with superforms. In section 2 we present the 0,1 and 2-forms while in section 3 we discuss the 3-form and the various complications it brings with it. In section 4 we first review the  ectoplasmic integration formula and then work out the various action formulas from 4,5 and 6-forms. We conclude by commenting on possible future research.  

\begin{center} {\bf {\large 1. HARMONIC SUPERSPACE}} \end{center}
{\bf The Coset Structure:} The harmonic superspace consists of the co-ordinates $\{x^m,\te^{\mu}_i,\tb^{\dot{\mu} i},y^{\dot{m}}\}$ where $m = 0...3$, $\mu,\md=0,1$, $i = 1,2$ and $\dot{m}=1,2$. Apart from the usual N=2 superspace co-ordinates therefore we now have two extra harmonic (spherical) co-ordinates $y^{\dot{m}}$ labelling $\frac{SU(2)}{U(1)} $. 

We define the coset in the usual way. Let $T^{++},T^{--}$ and $T^3$ be the generators of SU(2), with $T^3$ being the U(1) coset generator. Then any $g \in SU(2)$ can be written as 
$$g=e^{i[\lambda_{++}T^{++} +\lambda_{--}T^{--}]} e^{i\lambda_3 T^3}$$ 
so that any left coset is labelled by an element 
$$u=e^{i[\lambda_{++}T^{++} +\lambda_{--}T^{--}]}\in \frac{SU(2)}{U(1)}$$ Instead of using $\la^{\pm\pm}$ one can choose other co-ordinates $y^{\dot{m}}$ to paramatrize the coset space. 
$$
\la^{\pm\pm}=\la^{\pm\pm}(y^{\dot{m}})
$$
The vielbein\footnote{The vielbeins  $e_A = e_A^{\ M}e_M=e_A^{\ M}\p_M$ in purely bosonic manifolds  form an ``orthonormal'' set of tangent basis vectors, $e_A^{\ M}$ being the matrix of transformation. However, in fermionic spaces the concept of  a  metric is ill defined, and as  a  result $e_A$ can be thought of as  a  linearly independent set of tangent basis vectors, which we can of course choose according to our convenience.} for the coset space can be obtained applying Cartan's method (which was later independently discovered by physicists \cite{cartan}):
 $$u^{-1}\p_{\dot{m}}u=e_{\dot{m}}^{\ \dot{a}}T_{\dot{a}} + \om_{\dot{m}}^{\ 3} T_3$$
where $e_{\dot{m}}^{\ \dot{a}}$ are now the inverse vielbein in the coset space and $\dot{a}={++,--}$. It is however not necessary for us to compute the vielbein functions explicitly, its form in terms of the fundamental representation of an SU(2) matrix is sufficient. If we denote the fundamental SU(2) matrix  and its inverse respectively by
\begin{equation}
u_i^{\ u}=\left( \begin{array}{cc} u_1^{\ +}&u_1^{\ -}\\
                                               u_2^{\ +} &u_2^{\ -}
\end{array} \right), \ u_u^{\ i}=\left( \begin{array}{cc} u_+^{\ 1} & u_+^{\ 2}\\
                                                                           u_-^{\ 1}  & u_-^{\ 2} 
\end{array} \right)
\end{equation}
then 
\begin{equation}
e_{\dot{m}}^{\ \mp\mp}=u_{\pm}^{\ i}\p_{\dot{m}}u_i^{\ \mp}
\end{equation}

We can now write down  a  vielbein matrix for the ``flat'' harmonic superspace as
\begin{equation}
e_A^{\ M}=\left( \begin{array}{cccc}
\delta_a^m & 0& 0&0\\
i\s^m_{\al\ad}\tb^{\ad i}u_i^{\ u} & u_i^{\ u}\delta_{\al}^{\mu} &0 &0\\
-i\s^m_{\al\ad}\te^{\al i}u_u^{\ i} & 0 & -u_u^{\ i}\delta_{\ad}^{\dot{\mu}} &0\\
0&0&0&e_{\dot{a}}^{\ \dot{m}}
\end{array} \right)
\end{equation}
Here $_M=\{_m,_{\mu}^i,_{\md j},_{\dot{m}}\}$ and $ _A=\{_a,_{\al}^u,_{\ad v},_{\dot{a}}\}$.
 The fact that space-time is ``flat'' dictates that $e_a^{\ m}=\de_a^m$. The vielbein in the harmonic sector comes from the SU(2) group vielbein  \cite{cartan}, which we have obtained using Cartan's method. In the fermionic sector  we have  a  ``freedom'' and  with  this special choice the forms $\om^A=dx^M e_M^{\ A}$ (and in particular $\om^{\al}$'s) become supersymmetric, which in turn is because the spinor covariant derivatives  (Note: $e_A \equiv e_A^{\ M}\p_M$)   
$$D_{\al}^v\equiv e_{\al}^v=u_i^v\left(\frac{\p}{\p\te^{\al}_i}+i\s^m_{\al,\ad}\tb^{\ad i}\p_m\right),\ \ \D_{\ad v}\equiv e_{\ad v}=-u_v^i\left(\frac{\p}{\p\tb^{\ad i}}+i\s^m_{\al,\ad}\te^{\al}_i\p_m\right)$$ anti-commute with the supersymmetry charges
$$Q_{\al}^i=\left(\frac{\p}{\p\te^{\al}_i}-i\s^m_{\al,\ad}\tb^{\ad i}\p_m\right), \ \ 
Q_{\ad i}=\left(\frac{\p}{\p\te^{\ad i}}-i\s^m_{\al,\ad}\tb^{\al}_i\p_m\right)$$
{\bf Superforms:} Differential forms in ordinary (bosonic) manifolds are antisymmetric covariant tensors  \cite{forms}.
If $\omega^M=dx^M$ are the basis cotangent vectors then  a  p-form is given by 
$$ A_p=\frac{1}{p!} \omega^{M_1}\wedge\omega^{M_2}...\omega^{M_p}A_{M_p,...M_1}$$ 
where $\wedge$ is the antisymmetric wedge product  \cite{forms}. 
They  can be generalized to ``superforms'' in superspaces by  a  few  natural modifications  \cite{superforms,siegel,wess} of rules involving products of forms and the operation of $d$, both of which are important for our superform calculations. 
In superspace  we now have
$$\omega^{M}\wedge\omega^{N}=(-1)^{m+n}\omega^{N}\wedge\omega^{M}$$ where $m$ = 0 or 1 according to $M$ being bosonic or fermionic respectively. 
The $d$ operator as before takes p-forms to p+1-forms and is defined as follows:
$$ df=dx^M \p_Mf =\omega^M \p_Mf$$
$$d(\Omega_p\wedge\Phi_q) = \Omega\wedge d\Phi + (-1)^qd\Omega\wedge\Phi$$
\begin{equation}
d^2=0
\end{equation} 
Clearly $d\om^M=0$ but since we will work in flat basis   we need  $d\om^A$'s (which are non-zero), which can be computed from the vielbein matrix (3). However, it turns out that working with the operator $d$ is complicated and often requires working with explicit co-ordinates\footnote{There are   complicated connection-like terms present in $d\om^A$'s and it is also cumbersome to work with the operators $e_{\pm\pm}$.}. Fortunately we can replace $d$ with the covariant operator $D$ \cite{harm}:
$$d=\omega^Ae_A=(\omega^a\p_a+\omega^{\al}_uD_{\al}^u +\omega^{\ad u}\D_{\ad u}+\omega^{++}e_{++} +\omega^{--}e_{--}) $$
$$\longrightarrow (\omega^a\p_a+\omega^{\al}_uD_{\al}^u +\omega^{\ad u}\D_{\ad u}+\omega^{++}D_{++} +\omega^{--}D_{--})\equiv \omega^AD_A\equiv D$$
where we introduce the  covariant derivatives $D_{++},D_{--}$ and $D_0$ as
\begin{equation}
D_{\pm\pm}=u_i^{\pm}\frac{\p}{\p u_i^{\mp}}, \ D_0=u_i^{\pm}\frac{\p}{\p u_i^{\pm}}-u_i^{\mp}\frac{\p}{\p u_i^{\mp}}
\end{equation}
It is    possible to replace $d$ with $D$  (which differ from each other by connection terms \cite{harm}) because the action of both $d$ and $D$ on differential forms, which carry no harmonic index (and hence the connection terms do not contribute), is the same.
  
We obtain
$$D\omega^{c}=2i\omega^{\al}_u\wedge\omega^{\bd u}\s^c_{\al\bd}$$
\begin{equation}
D\omega_v^{\bb}=-\omega^{\bb}_-\wedge\omega^{--}\delta_v^+-\omega^{\bb}_+\wedge\omega^{++}\delta_v^-
\end{equation}
With the understanding that $d$ and $D$ can be used interchangebly we, in the rest of our paper, will denote either of them by $d$.  

One can write down the covariant operator in terms of $\frac{\p}{\p y^{\dot{m}}}$, but it is well known that functions on the coset space have definite charge ($D_0f^q=qf^q$) and can be expanded as a harmonic series of fundamental functions constructed out of product of $u$'s  \cite{harm}:
$$f^q(x,\te,y)=\sum_n u_{(i_1}^+...u_{i_{n+q}}^+u_{j_1}^-..u_{j_n)}^- f^{i_1i_2...j_n}(x,\te)$$ 
Hence it is convenient to work with partial derivatives of $u$'s\footnote{ Clearly not all $u_i^u$'s are independent, but partial derivatives with respect to them is perfectly well defined. Even though one cannot express $\frac{\p}{\p u_i^{\pm}}$ in terms of $\frac{\p}{\p y^{\dot{m}}}$ directly, one can  verify  using $\frac{\p}{\p y^{\dot{m}}}=\frac{\p u_i^{u}}{\p y^{\dot{m}}}\frac{\p}{\p u_i^{u}}$  that the covariant derivatives (6)  indeed can be expressed completely in terms of the $y$'s.}
 rather than $y$'s.
 The various (anti)commutators involving the covariant derivatives needed for later calculations are as follows:
$$\{D_{\al}^u,D_{\bb}^v\}=\{\D_{\ad}^u,\D_{\bd}^v\}=[\p_a,D_B]=0$$
$$\{D_{\al}^u,\D_{\bd}^v\}=-2i\s^c_{\al\bd}\delta_u^v\p_c$$
$$[D_{\al}^u,D_{\mp\mp}]=-D_{\al}^{\pm}\delta_{\mp}^u$$
$$[D_{--},D_{++}]=D_0=q, \ [D_0,D_{\pm\pm}]=\mp D_{\pm}$$
{\bf The Star Conjugation:} In order to obtain irreducible supermultiplets it should be sufficient to consider only ``real'' multiplets (since supersymmetry respects reality), but an arbitrary  harmonic function with charge q is necessarily complex; their complex conjugates have charge $-$q. Hence the need to define a new reality operation. Indeed, if q is even, i.e. if the isospins in the harmonic expansion take integer values, one can arrange for the tensor co-efficients to be real. To this end one has to accompany the complex conjugation by a new involution  \cite{harm}, an antipodal map (on the sphere) that acts only  on the harmonic part. Such an involution ($\diamondsuit$) indeed exists, and under it the harmonic functions $u_i^u$ transform as $u_i^{\pm}\stackrel{\diamondsuit}{\rightarrow} \pm u_i^{\mp}$. It follows that the operation squares to $-1$ on the harmonics, as well as any odd product of harmonics, and to 1 on even products. We  define  a  new generalized conjugation $(\ )^{\x}$ as the product of ordinary complex conjugation and the antipodal map: $(\ )^{\x}=\overline{(\ )}^{\diamondsuit}$. One has to now carefully define/compute how the reality operation acts on fermionic functions and forms (see appendix A). The rules that we use for computations are summarized below:
$$(f_M f_N)^{\x}=(-)^{mn}f_M^{\x}f_N^{\x}$$
$$(\Omega^M f_M)^{\x}=(-)^m \Omega^{M\x}f_M^{\x}$$
$$(\Omega^M\wedge\Lambda^N)^{\x}=(-)^{mn}\Omega^{M\x}\wedge\Lambda^{N\x}$$
$$\om^{\al\x}_{\pm}=\mp\om^{\ad \mp},\ \om^{\ad \mp\x}=\pm\om^{\al}_{\pm}$$
$$D_{\al}^{\pm\x}=\left\{ \begin{array}{c} \pm\D_{\ad \mp}$ acting on bosons$\\
 \ \ \mp\D_{\ad \mp}$ acting on fermions$ \end{array}\right.$$
$$\om^{\pm\pm\x}=\om^{\pm\pm}$$
\begin{equation}
D_{\pm\pm}^{\x}=D_{\pm\pm}
\end{equation}

\begin{center} {\bf {\large 2. SUPERMULTIPLETS FROM SUPERFORMS}} \end{center}
{\bf The Gauge Structure:} Superforms have an inbuilt gauge structure. In general consider  a  p-form (gauge field) $A_p$. We define its field strength $F$ as
\begin{equation}
 F_{p+1}=dA_p
\end{equation}
Then $F$ is invariant under the  gauge transformation\footnote{As is clear,  this is an Abelian gauge transformation, and for the rest of the paper we will concern ourselves with abelian gauge forms only. }
\begin{equation}
\de A_p=d\Lambda_{p-1}
\end{equation}
and satisfies the Bianchi identity
\begin{equation}
B_{p+2}\equiv dF_{p+1}=0
\end{equation}

Thus now one can construct gauge invariant actions for the dynamic fields $A$ using their field strength $F$; and we have produced gauge theories. This is the whole story when we consider only bosonic manifolds, but in superspace things become more complicated. The gauge form $A$  contains not only gauge fields but also matter fields, which  (in definite combinations) constitute what are known as multiplets, and therefore we are now looking at theories for gauge supermultiplets. Moreover, in general the gauge field $A$ is highly reducible (i.e. will contain several multiplets), and we have to constrain them to get to  a  single irreducible supermultiplet (i.e. the field content in $A$ will be that of  a  single  multiplet), and this is where the complications come in. There is no well defined algorithm for identifying the ``right'' constraints. However, we do know that these constraints have to be supersymmetric and gauge invariant, and hence a natural way  to impose  the constraints is by putting some of the field strength components to zero\footnote{Notice that the field strength components are not only gauge invariant but in our choice of flat basis are also supersymmetric.}  \cite{siegel,wess}. It is also advantageous to remove the unphysical gauge degrees of freedom from $A$, so that finally what we are left with  can be expressed in terms of  a  single  ``prepotential'' ($V$) having the field content  of an irreducible supermultiplet. Often we also have  a  ``residual'' gauge freedom which translates into  a  gauge transformation for $V$. This procedure was successfully applied to N=1 superforms and one could obtain all the known important N=1 supermultiplets by this method \cite{n=1,siegel}. 

In the solutions of the superforms one however observes  a  curious fact. The structure of any p-form, whether  it is the gauge field, field strength, Bianchi identity or the gauge parameter, remains the same! For example, when 
one computes the field strength of 1-form it can be expressed in terms of  a  chiral spinorial ``field function'' $W_{\al}$ (which is of course expressible in terms of the prepotential of the 1-form). If one  constrains the 2-form gauge potential, one finds that it can be also written in terms of  a  chiral spinorial prepotential $V_{\al}$ and moreover in exactly the same manner. This  trend continues through out the ``form-table'' and facilitates  calculation greatly apart from adding beauty to the structure. For completeness we enumerate the table for N=1 below  \cite{n=1,siegel}.
$$\begin{array}{lll} p=0:&\ A=\2 (V+\overline{V})& W=i(\overline{V}-V)\\
p=1: & \ A_{\al}=i\frac{1}{2}D_{\al}V, \ A_{\al\ad}=\frac{1}{2}[\D_{\ad},D_{\al}]V &W^{\al}=i\D^2D^{\al}V\\
p=2: & \ A_{\al\bb \bd}=-\epsilon_{\al\bb}V_{\bd}& 
 W=\2(D_{\al}V^{\al}+\D_{\ad}V^{\ad})\\
&\ A_{\al\ad\bb\bd}=\frac{1}{2}(i\epsilon_{\ad\bd}D_{(\al}V_{\bb)}+c.c.)&\\
p=3:& \ A_{\al\bd\ga\gd}=V, \ A_{\al\bb\bd\ga\gd}=\e_{\bd\gd}\e_{\al(\bb}D_{\ga)}V & W=\D^2V\\
&
 \ A_{abc}=\e_{abcd}\s^d_{\delta\dot{\delta}}[\D^{\dot{\delta}},D^{\delta}]V & \\
p=4:& \ A_{\al\bb\ga\gd\delta\dot{\delta}}=-2i\e_{\gd\dot{\delta}}\e_{\al(\ga}\e_{\delta)\bb}V & W=0\\
 &\ A_{\al bcd}=2\e_{abcd}\s^a_{\al\ad}\D^{\ad}V& \\
 & \ A_{abcd}=2i\e_{abcd}(\D^2\overline{V}
- D^2V)&  \end{array}$$
For odd p, $V=\overline{V}$, and for even p, $\D_{\ad}V=0$. Also $A_{\al\ad}=\s^a_{\al\ad}A_a$.
\vspace{5mm}
\\
{\bf Lower Harmonic Superforms:} We want to now perform  a  similar analysis for the harmonic superspace. It turns out that the 0,1 and 2 forms describe the scalar ($\om$-hypermultiplet), the vector (Yang-Mills) and the tensor multiplets respectively. When we reach the 3-form we encounter additional complications which will be discussed in the next section.

Before we begin, we should perhaps clarify our notation for the component indices, ${}_A=\{ {}_{\pm\pm},{}_{\al}^{\pm},{}_{\ad\pm},{}_a\}$. Note that everything appears as a lower index except the $\pm$ associated with the $\al,\bb$s. Hence, whenever $\pm$ indices appear as an upper index they are to be associated  with the $\al,\bb$ type indices appearing in the lower index in a sequential manner.  

{\bf 0-form:} $A_0$ is just  a  function which also becomes the prepotential.
\begin{equation}
V=A
\end{equation}
Imposing reality implies $V=V^{\x}$. Then the field strengths are given by
\begin{equation}
F=\om^A F_A=dA=\om^A D_A A=\om^A D_A V
\end{equation}
 We constrain $V$  by imposing
$$ F_{\al}^+=0$$
\begin{equation}
\Ra D_{\al}^+V=0
\end{equation}
and thus $V$ is analytic ($D_{\al}^+V=\D_{\ad -}V=0$)
Then the field function  is given by
\begin{equation}
F_{--}=\Dp V\equiv W^{++}
\end{equation}
All other field components can be now expressed in terms of the field function $W^{++}$, either implicitly or explicitly. For example we have
$$D_{\al}^-W^{++}=D_{\al}^-\Dp V=\Dp D_{\al}^-V=\Dp F_{\al}^-$$ 
(since $D_{\al}^+V=0$), and thus $F_{\al}^-$ is implicitly\footnote{One can check that differential equations of the form $\Dp B^q=C^{q+2}$ uniquely determine $B^q$ in terms of $C^{q+2}$ if the charge q is negative, or up to an arbitrary function $u_{i_1}^+...u_{i_q}^+B^{i_1...i_q}(X)$ if q is positive.}
given in terms of $W^{++}$ by
\begin{equation}
\Dp F_{\al}^-=D_{\al}^- W^{++}
\end{equation}

Similarly $F_{\ad +}$ is given by
\begin{equation}
\Dp F_{\ad +}=\D_{\ad +}W^{++}
\end{equation}
Further we have
$$F_{\al\ad}=i\2 \{D_{\al}^+,\Dap\}V=i\2 D_{\al}^+,\Dap V$$
\begin{equation}
F_{\al\ad}=i\2 D_{\al}^+ F_{\ad +}
\end{equation}
and finally we have
$$ [\Dp,\Dm]V=0=\Dp F_{++}-\Dm F_{--}$$
\begin{equation}
\Dp F_{++}-\Dm W^{++}=0
\end{equation}
Thus the entire field strength is determined by  a  single field function $W^{++}$. It is  a  trivial exercise  to check that $W^{++}$ is real and analytic 
in terms of which we can now write down the action for V: 
\begin{equation}
S=-\2\int d^2y \ d^4x D^{-4}(W^{++})^2
\end{equation}
We recognize that this is the well known free action for the $\om$-hypermultiplet \cite{harm}.

{\bf 1-form:} This has been discussed quite extensively,  though in  a  slightly different framework \cite{harm}. We here start with $A_1=\om^A A_A$. The master formula for computing $F$ is then
\begin{equation}
F=dA=\om^A\wedge\om^B D_B A_A + d\om^A A_A
\end{equation}
We can obtain the components of $F$ using (20) and then equating the linearly independent wedge products. We put constraints
$$F_{\bb\al}^{++}=0=D_{(\bb}^+A_{\al)}^+ \Ra A_{\al}^+=D_{\al}^+K$$
where $K$ is an arbitrary function, which can be eliminated  using gauge invariance
$$\de A_{\al}^+=D_{\al}^+\Lambda$$
In this gauge $A_{--}$ becomes the prepotential. 
\begin{equation}
V^{++}\equiv A_{--}
\end{equation}
Reality on $A$ then implies $V^{++\x}=V^{++}$ and we constrain it by imposing 
$$ F^{\ +}_{--\al}=\Dp A_{\al}^+ -D_{\al}^+A_{--}=-D_{\al}^+V^{++}=0$$
Thus $V^{++}$ is analytic. We further impose 
\begin{equation}
F_{++--}=0=\Dm V^{++}-\Dp A_{++}
\end{equation}
which gives us $A_{++}$ in terms of $V^{++}$. Note the similarity between (22) and (18). Imposing constraints $F^{++\ -}_{\ \al}=0$ we obtain
\begin{equation}
\Dp A_{\al}^-=D_{\al}^- V^{++}
\end{equation} again same as (15). Similarly $F_{\ad -\al}^{\ \ -}=0$ determines
\begin{equation}
A_{\al\ad}=i\2 \Dam A_{\al}^-
\end{equation}
like (17). It is easy to check that the first\footnote{By first we mean the lowest dimension.} nonzero field strength component appears in $F_{\bb\al}^{-+}$:
$$F_{\bb\al}^{-+}= D_{\bb}^-A_{\al}^+ +D_{\al}^+A_{\bb}^-=-D_{\al}^+D_{\bb}^+A_{++}=\2 \e_{\al\bb}D^{+2}A_{++}$$ or
\begin{equation}
F_{\bb\al}^{-+}=\e_{\al\bb} W
\end{equation} where we define the field function $W$ as
\begin{equation}
W=-\2 D^{+2}A_{++}
\end{equation}
Using Bianchi identities one can compute all the other non-zero components of the field strength in terms of $W$. One finds
\begin{equation}
F_{\bb \al\ad}^{\pm}=\mp i\2 \e_{\al\bb}\overline{D}_{\ad\mp} W
\end{equation}
\begin{equation}
F_{\bb\bd\al\ad}=-\8(\e_{\al\bb}\overline{D}_{(\bd -}\overline{D}_{\ad)+}W -\e_{\ad\bd}D_{(\bb}^-D_{\al)}^+ W^{\x})
\end{equation}
It is not difficult to check that $W$ is chiral
\begin{equation}
D_{\al}^{\pm}W=0
\end{equation}
and further it is independent of $u$:
\begin{equation}
D_{\pm\pm}W=0
\end{equation}
One can now write down the action for the Yang-Mills multiplet $V^{++}$ in terms of the field function $W$ as
\begin{equation}
S_{YM}=\4\int d^4x D^4 W^2
\end{equation}

{\bf 2-form:} We start now with the 2-form $A_2=\2\om^A\om^B A_{BA}$. The master formula for computing $F$ again follows from (4)
\begin{equation}
F=dA=\frac{1}{6}\om^A\om^B\om^C F_{CBA}=\2\om^A\om^B\om^C D_CA_{BA} + \om^Ad\om^BA_{BA}
\end{equation}
As before one can compute the individual components by equating the independent wedge products.
As for the  1-form we impose constraints and use gauge invariance to obtain our irreducible multiplet. Imposing 
$$F_{\ga\bb\al}^{+++}=0 \Ra A_{\bb\al}^{++}=D_{(\bb}^+K_{\al)}^+$$ where $K_{\al}^+$ is an arbitrary function which can be eliminated  using $\La_{\al}^+$
$$\de A_{\bb\al}^{++}= D_{(\bb}^+\La _{\al)}^+$$ 
Continuing in this fashion we can put $A_{--\al}^{+}=0$ imposing $F^{++}_{--\bb\al}=0$ and using $\La_{--}$. 
Further, $\de A_{++--}=-\Dp \La_{++} +...$ and hence 
\begin{equation}
A_{++--}=V'
\end{equation}
where $V'$ is independent of $u$, and it will soon turn out to be an irrelevant constant prepotential. Now $\de A_{++\al}^{+}=... -\La_{\al}^-$ and thus using it\footnote{One has to be careful that  the chronological sequence of using the gauge parameters is such that the one used later does not spoil the gauge condition fixed by prior parameters.}  we can put $ A_{++\al}^{+}=0$. We continue to impose further constraints
$$F_{++--\al}^{+}=\Dm A_{--\al}^{+}-\Dm A_{++\al}^{+}+D_{\al}^+A_{++--}-A_{--\al}^{-}=0 $$
\begin{equation}
\Ra A_{--\al}^{-}=D_{\al}^+ V'
\end{equation}
Similarly $$F_{++--\al}^{-}=0 $$
\begin{equation}
\Ra\Dp A_{++\al}^{-}=2D_{\al}^-V'\sim u_i^-V_{\al}^{'i}(X)
\end{equation}
whereas $$A_{++\al}^{-}=u_i^-u_j^-u_k^- A_{\al}^{ijk}(X) +...$$
where X denotes co-ordinates other than the harmonic ones.
Hence equating the harmonic functions it is clear that both the right-hand side and the left-hand side of (35) have to be zero. Now since $D_{\al}^-V'(X)=0$ we get $D_{\al}^+V'(X)=0$ which implies not only 
$$ D_{\al}^i V'=0\Ra V'=constant$$ but also 
$$ A_{++\al}^{\ -}=0$$ 
Thus we finally have
$$A_{++--}= A_{\dot{b}\al}^{\ u}=0$$ 
Now
$$ F_{++\bb\al}^{++}=0=\Dm A_{\bb\al}^{++} -D_{(\bb}^+A_{++\al)}^{+}-A_{(\bb\al)}^{-+} $$
\begin{equation}
\Ra A_{\bb\al}^{-+}=\e_{\bb\al}V
\end{equation}
and we have found our ``true'' prepotential in  $V$. Our next task is to find all other non-zero components of $A$ and express them in terms of $V$: 
\begin{equation}
F_{--\bb\al}^{--}=0=\Dp A_{\bb\al}^{--} -D_{(\bb}^-A_{--\al)}^{-}-A_{(\bb\al)}^{-+}\Ra \Dp A_{\bb\al}^{--}=0\Ra A_{\bb\al}^{--}=0
\end{equation}
We observe $\de A_{\bd -\al}^-=... + 2i \La_{\al\bd}$ and thus we can put it to zero, too. This together with the constraints 
$$F_{--\bd+\al}^{\ -}=F_{++\bd-\al}^{\ +}=0\Ra A_{\bd\pm\al}^{\ \ \ \pm}=0$$ 
Finally imposing 
$$F_{--\bd-\al}^{\ -}=F_{++\bd-\al}^{\ -}=0\Ra A_{\pm\pm a}=0$$
Thus except for our prepotential, all other dimension 1 and lower components of $A$ are zero. To impose constraints on our prepotential $V$ we impose 
$$F_{\ga\bb\al}^{-++}=F_{\ga\bb\al}^{--+}=0 \Ra D_{\al}^{\pm}V=0$$ 
or in other words $V$ is chiral, just as the field function of the 1-form. Imposing $$F^{-+}_{++\bb\al}=0 \Ra \Dp V=0$$
 i.e. $V$ is $u$ independent. All the other components of $A$ can now be obtained in terms of $V$ by imposing further constraints through field strength components. Thus $F_{\gd -\bb\al}^{\ -+}=F_{\gd +\bb\al}^{\ -+}=F_{(\gd-\ga,\al)\ad}^{\ -}=0$ gives us
\begin{equation}
A_{\bb\al\ad}^{\pm}=\pm i\2\e_{\bb\al} \overline{D}_{\ad\mp}V
\end{equation}
and
\begin{equation}
A_{\bb\bd\al\ad}=-\8(\e_{\al\bb}\overline{D}_{(\bd -}\overline{D}_{\ad)+}V -\e_{\ad\bd}D_{(\bb}^-D_{\al)}^+ V^{\x})
\end{equation}
 Obviously  these are reminiscent of (27) and (28). The first nonzero component of the field strength occurs in the completely anti-symmetric part of $F_{\gd-\ga\al\ad}^{\ -}$ One obtains after some straightforward manipulations
\begin{equation}
F_{\gd-\ga\al\ad}^{\ -}=\e_{\ga\al}\e_{\gd\ad}W
\end{equation}
where the field function $W$ is given by
\begin{equation}
W=i\4 (\overline{D}_-\overline{D}_+V - D^-D^+V^{\x})
\end{equation}
Clearly $W^{\x}=W$, i.e. the field function is real. Using Bianchi identities one obtains other non-zero components of the field strength in terms of $W$ as expected. We find
\begin{equation}
F_{\gd-\ga\al\ad}^{\ +}=\e_{\ga\al}\e_{\gd\ad}W^{++}
\end{equation}
while
\begin{equation}
F_{\gd+\ga\al\ad}^{\ -}=\e_{\ga\al}\e_{\gd\ad}W^{--}
\end{equation}
where 
$$\Dp W=W^{++},\ \Dm W=W^{--}$$ Further from these definitions one finds
\begin{equation}
\Dp W^{++}=D_{\al}^+W^{++}=0
\end{equation}
i.e. $W^{++}$ is constrained analytic, and hence it is convenient to choose it as the field function rather than $W$:
\begin{equation}
\Dm W^{++}=2W=-\Dp W^{--}
\end{equation}
In terms of $W^{++}$ the components of the field strength read as
\begin{equation}
F_{\ga\bb\bd\al\ad}^{\pm}=-i\4\e_{\bd\ad}\e_{\ga(\al}D_{\bb)}^{\mp}W^{\pm\pm}
\end{equation}
\begin{equation}
F_{\de\dd}=\8(\Ddm D_{\al}^+W^{--}+\Ddp D_{\al}^-W^{++})
\end{equation}
where we define $F_{\de\dd}$ as 
\begin{equation}
F_{dcba}=\e_{dcba}F^d;\ F_{\de\dd}=\s_{d\de\dd}F^d
\end{equation}
One can now construct the action for the tensor multiplet $V$ in terms of the field function as\footnote{In the CP(1) harmonic superspace language, the action of the tensor multiplet is given in terms of  a  holomorphic function $W(t)$, as $S=\oint \frac{dt}{t}D^4 W(t)$ \cite{trans}. $W(t)$ is of course related to our $W^{++}$ while $t$ can be viewed as  a  special co-ordinate system spanning $\frac{SU(2)}{U(1)}$. For details see appendix B.}
\begin{equation}
S_{free}=\int d^2y\ d^4x \overline{D}_+^2D^{-2}W^{++}
\end{equation}

A careful look at the 2-form will reveal that the $u$-dependence of the form is trivial, i.e. completely determined; all that we have is  a  chiral prepotential $V(X)$, while the field function is given in terms of the functions $W^{ij}(X)$ ($W^{++}=u_i^+u_j^+W^{ij}$). Thus it is to be expected that  a  formulation of the 2-form will exist in the ordinary N=2 superspace and indeed it does \cite{ten}. With the above mentioned identification one can check our results with \cite{ten}.

\begin{center} {\bf {\large 2. THE 3-FORM }} \end{center}
{\bf Advent of Multi-Prepotentials:}
We have seen that like the N=1 case the 0,1 and 2 superforms of the N=2 theory follow  a  linear chain where the field strength of one is the prepotential of the next, and more importantly there is only  a  single independent real prepotential at each level. However, this simple structure breaks down, to some extent at least, when we go to 3 or higher forms. We start getting more than one  independent prepotential that are not related by gauge transformations. As  a  result it becomes very difficult to work with these multiple prepotentials simultaneously, with complicated differential equations that have to be solved. Clearly it is advisable then to somehow disentangle  the equations and work with  a  single prepotential at  a  time. Due to the linear nature of the theory ($F$ is linear in $A$), the most general solution will then be  a  linear combination of the independent solutions.
\begin{center}
\includegraphics[width=5.0in,height=6.0in]{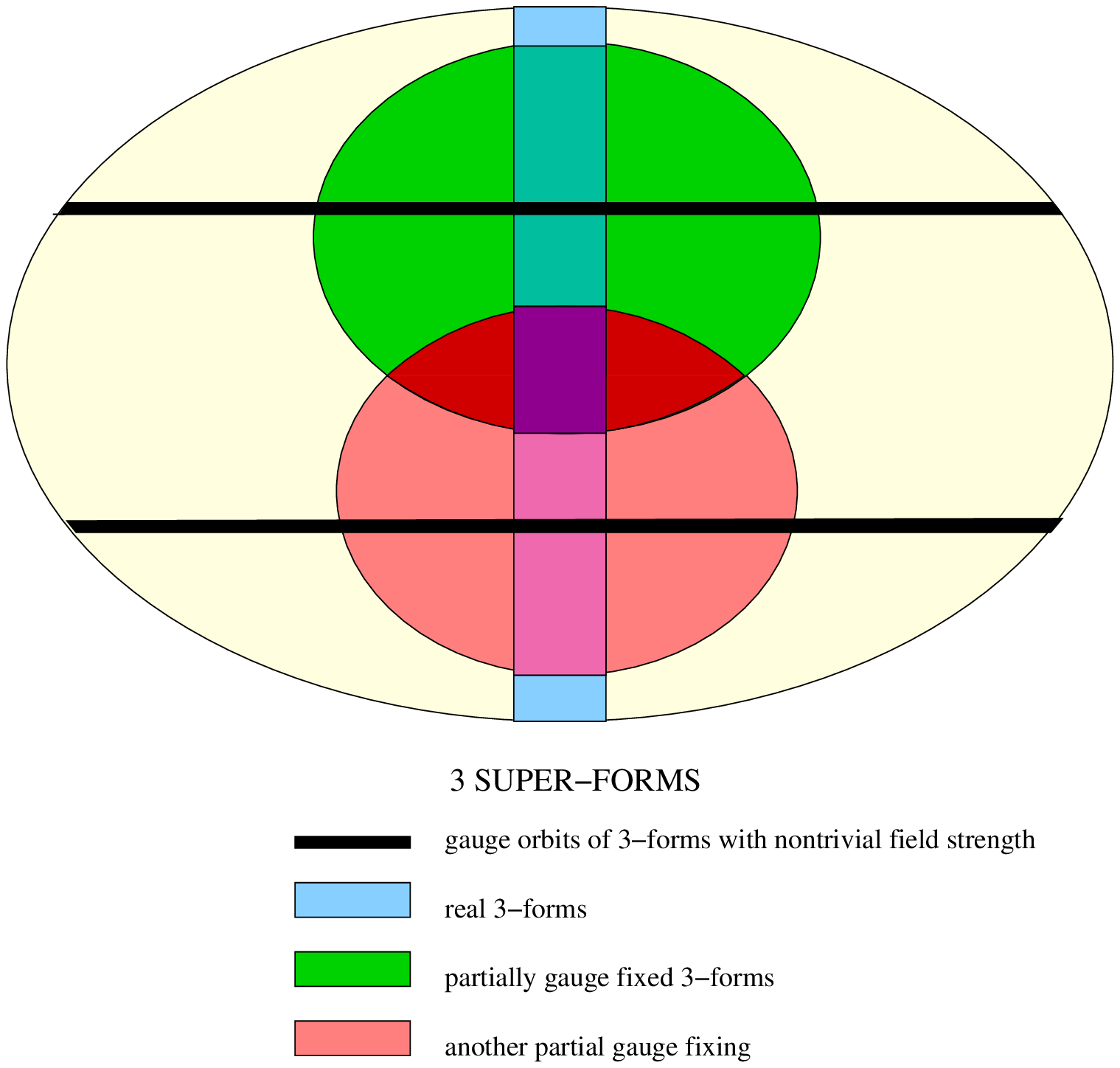}
\end{center}
To accomplish the same we devised an approach  based on ``complex gauge fixing''. We do  a  partial gauge fixing (complex) first and then impose the reality condition so that we end up with  a  single prepotential. If we do the initial complex gauge fixing in  a  slightly different way then we may end up with  a  different prepotential. The diagram (previous page) should make things clear:

By applying this method we keep things supersymmetric as well as preserve the gauge structure, and  ``voila!'', accomplish  a  disentaglement  of the prepotentials. The only defect of this approach is that even if we obtain all the independent prepotentials by this method it will be difficult to prove that we have obtained the most general solution of the problem. However, this is more of  a  technical (or mathematical) issue and that gives us an excuse to overlook it. In what follows below we present two independent prepotentials for the 3-form but we strongly suspect that there is  a  third one.
\vspace{5mm}
\\
{\bf The ``Chain'' 3-Form:} We start with  a  complex 3 form 
$$A_3=\frac{1}{6} \om^A\om^B\om^C A_{CBA}$$ The master formula for computing $F$ is given by
$$F=\frac{1}{24} \om^A\om^B\om^C\om^D F_{DCBA}=dA=\frac{1}{6}\om^A\om^B\om^C\om^D D_D A_{CBA}$$
\begin{equation}
 +\2\om^A\om^B d\om^CA_{CBA} 
\end{equation}
The procedure of imposing  the constraints and using the gauge freedom to put some components of $A$ to zero is very similar to what we used in 1 and 2-form analysis and hence we will skip the details. We obtain in chronological steps
$$F_{\de\ga\bb\al}^{++++}=0\Ra A_{\ga\bb\al}^{+++}=0,\ \ (\La_{\bb\al}^{++})$$
$$F_{\de\ga\bb\al}^{-+++}=0\Ra A_{\ga\bb\al}^{-++}=0,\ \ (\La_{\bb\al}^{-+})$$
$$F_{\de\ga\bb\al}^{--++}=0\Ra A_{\ga\bb\al}^{--+}=0,\ \ (\La_{\bb\al}^{--})$$
$$F_{--\ga\bb\al}^{+++}=0\Ra A_{--\bb\al}^{++}=0,\ \ (\La_{--\al}^{+})$$
$$F_{--\ga\bb\al}^{-++}=0\Ra A_{--\bb\al}^{-+}=0,\ \ (\La_{--\al}^{-})$$
$$F_{++\ga\bb\al}^{+++}=0\Ra A_{++\bb\al}^{++}=0,\ \ (\La_{++\al}^{+})$$
$$F_{++\ga\bb\al}^{-++}=0\Ra A_{++\bb\al}^{-+}=0,\ \ (\La_{++\al}^{-})$$
$$F_{++--\bb\al}^{++}=0\Ra A_{++--\al}^{+}=0,\ \ (\La_{++--})$$
$$A_{++--\ad+}=0,\ \ (\La_{++\ad-})$$
At this point we impose the reality condition on $A_3\Ra A_{++--\al}=0$ among other things. One can then check the following
$$F_{++--\bb\al}^{-+}=0\Ra A_{--\bb\al}^{--}=0$$
$$F_{++--\bb\al}^{--}=0\Ra A_{++\bb\al}^{--}=0$$
$$F_{++\ga\bb\al}^{--+}=0\Ra A_{\ga\bb\al}^{---}=0$$
So we have
$$A_{\dot{c}\bb\al}=A_{\ga\bb\al}=0$$
We now use some more gauge freedom
$$A^{-}_{++\bd+\al}=0,\ \ (\La_{\bd+\al}^-)$$
$$F_{\dd-\ga\bb\al}^{\ \ +++}=0\Ra A_{\dd-\bb\al}^{\ \ ++}=0, \ \ (\La_{\bd-\al}^{\ \ +})$$
$$A^{\ +}_{++\bd-\al}=0,\ \ (\La_{\bd+\al}^+)$$
$$A^{\ +}_{++\bd+\al}=0,\ \ (\La_{++\bd\al})$$
$$A^{\ +}_{--\bd+\al}=0,\ \ (\La_{--\bd\al})$$
and impose further constraints
$$F_{++--\bd-\al}^{\ \ +}=0\Ra A^{\ +}_{--\bd-\al}=0$$
$$F_{++--\bd+\al}^{\ \ +}=0\Ra A^{\ +}_{++\bd-\al}+A^{\ -}_{--\bd+\al} + 2iA_{++--\al\bd}=0$$
Taking the star conjugate (equivalent to looking at $F_{++--\bd-\al}^{\ \ -}$) we have
$$A^{\ +}_{++\bd-\al}+A^{\ -}_{--\bd+\al} - 2iA_{++--\al\bd}=0$$
and thus 
$$A_{++--\al\bd}=A^{\ -}_{--\bd+\al}=0$$
We now have 
$$A_{++--a}=A_{\dot{c}\bd\al}=0$$
Next 
$$A_{-- ba}=0, \ \ (\La_{ba})$$
and $$F_{++--ba}=0\Ra A_{++ba}=0\ra A_{\dot{c}ba}=0$$
Last but not least we use another set of gauge freedom
$$A_{\dd-\bb\al}^{\ \ -+}=0,\ \ (\La_{\bb a}^+)$$
$$A_{\dd+\bb\al}^{\ \ -+}=0,\ \ (\La_{\bb a}^-)$$
$$F_{++\gd-\bb\al}^{\ \ ++}=0\Ra A_{\dd+\bb\al}^{\ \ ++}=0$$
$$F_{++\gd+\bb\al}^{\ \ ++}=0\Ra A_{++\bb a}^{+}=0$$
$$F_{++\gd-\bb\al}^{\ \ -+}=0\Ra A_{\dd-\bb\al}^{\ \ --}=0$$
$$F_{--\gd+\bb\al}^{\ \ --}=0\Ra A_{\dd+\bb\al}^{\ \ --}=0$$
$$F_{--\gd-\bb\al}^{\ \ -+}=0\Ra A_{--\bb a}^{-}=0$$
$$F_{++--\bb a}^{-}=0\Ra A_{++\bb a}^{-}=0$$
$$F_{++--\bb a}^{+}=0\Ra A_{--\bb a}^{+}=0$$
$$\Ra A_{\dot{c}\bb\al}=A_{\gd\bb\al}=0$$ and thus all components of $A$ which have dimension less than 2 are zero.
Finally 
$$F_{\dd+\ga\bb\al}^{\ \ --+}=0\Ra A_{\ga\bb a}^{--}=0$$
$$F_{\dd-\ga\bb\al}^{\ \ -++}=0\Ra A_{\ga\bb a}^{++}=0$$
$$F_{\dd-\ga\bb\al}^{\ \ --+}=F_{--\ga\bb a}^{--}=0\Ra A_{\ga\bb a}^{-+}=0$$
i.e. $$A_{\ga\bb a}=0$$
Now
$$
F_{\dd-\gd+\bb\al}^{\ \ \ \ --}=0\Ra A_{(\gd+\bb\al\ad)}^{\ \ -}=A_{\gd+(\bb\al)\ad}^{\ \ -}=0$$
Thus the prepotential of the 3-form appears as 
\begin{equation}
A_{\gd+\bb\al\ad}^{\ \ -}=\e_{\gd\ad}\e_{\bb\al}V^{--}
\end{equation}
Similarly
$$F_{\dd-\gd+\bb\al}^{\ \ \ \ ++}=0\Ra$$
\begin{equation}
A_{\gd-\bb\al\ad}^{\ \ +}=\e_{\gd\ad}\e_{\bb\al}V^{++}
\end{equation}
Now 
\begin{equation}
F_{\dd-\gd+\bb\al}^{\ \ \ \ -+}=0\Ra A_{\dd-\bb\al\gd}^{\ \ -}+A_{\gd+\al\bb\dd}^{\ \ +}=0
\end{equation}
and 
\begin{equation}
F_{\dd-\gd-\bb\al}^{\ \ \ \ --}=0\Ra A_{(\dd-(\bb\al)\gd)}^{\ \ -}=0
\end{equation} 
i.e. the totally symmetric part of $A_{\dd-\bb\al\gd}^{\ \ -}$ is zero. Then the constraint $F_{--\gd+(\bb\al)\ad}^{\ \ -} = 0$ together with (54) implies that $A_{\dd-\bb\al\gd}^{\ \ -}$ is totally antisymmetric:
\begin{equation}
A_{\gd-\bb\al\ad}^{\ \ -}=\e_{\gd\ad}\e_{\bb\al}V
\end{equation}
\begin{equation}
A_{\gd+\bb\al\ad}^{\ \ +}=\e_{\gd\ad}\e_{\bb\al}V'
\end{equation}
The reality condition implies all $V$'s are real and further
\begin{equation}
V+V'=0
\end{equation}
A priori therefore we have three independent $V$'s, and here comes  a  tricky issue:  We have to put constraints to relate them, but there is an ambiguity as to which constraints to impose. We here first  list the set of relevant constraints: 
\begin{equation}
F_{--\gd+\bb\al\ad}^{\ \ +}=0\Ra \Dp V'+V^{++}=0
\end{equation}
\begin{equation}
F_{++\gd-\bb\al\ad}^{\ \ -}=0\Ra \Dm V+V^{--}=0
\end{equation}
\begin{equation}
F_{--\gd+\bb\al\ad}^{\ \ -}=0\Ra \Dp V^{--}+V-V'=0
\end{equation}
\begin{equation}
F_{++\gd-\bb\al\ad}^{\ \ +}=0\Ra \Dm V^{++}+V'-V =0
\end{equation}
\begin{equation}
F_{--\gd-\bb\al\ad}^{\ \ +}=0\Ra \Dp V^{++}=0
\end{equation}
\begin{equation}
F_{++\gd+\bb\al\ad}^{\ \ -}=0\Ra \Dm V^{--}=0
\end{equation}
One can work out that if we impose all of these above constraints, we indeed get  a  consistent solution, but the field strength vanishes, i.e. the multiplet becomes ``trivial''. If we want to find  a  nontrivial multiplet then we need to select the constraints carefully. We can get some hints from studying the field strength of the 2-form. Equation (45) relating the field functions seems to suggest that we impose the constraints (60) and (61) so that we end up with 
\begin{equation}
\Dm V^{++}=2V=-\Dp V^{--}
\end{equation}
which is identical to (45). As we will see it turns out to be the right choice. The field functions are then given by the constraints that we don't impose.
\begin{equation}
F_{--\gd+\bb\al\ad}^{\ \ +}=\e_{\bb\al}\e_{\gd\ad}(\Dp V'+V^{++})\equiv \e_{\bb\al}\e_{\gd\ad}W^{++}
\end{equation}
\begin{equation}
F_{--\gd-\bb\al\ad}^{\ \ -}=\e_{\bb\al}\e_{\gd\ad}(\Dm V+V^{--})\equiv \e_{\bb\al}\e_{\gd\ad}W^{--}
\end{equation}
\begin{equation}
F_{--\gd-\bb\al\ad}^{\ \ +}=\e_{\bb\al}\e_{\gd\ad}(\Dp V^{++})\equiv \e_{\bb\al}\e_{\gd\ad}W^{4+}
\end{equation}
\begin{equation}
F_{++\gd+\bb\al\ad}^{\ \ -}=\e_{\bb\al}\e_{\gd\ad}(\Dm V^{--})\equiv \e_{\bb\al}\e_{\gd\ad}W^{4-}
\end{equation}
However, these four field functions are not all independent, and we can for example determine everything in terms of $W^{4+}$ through Bianchi identities: 
\begin{equation}
B_{++--\gd-\bb a}^{\ \ +}=0\Ra W^{++}=-\2 \Dm W^{4+}
\end{equation}
\begin{equation}
B_{++--\gd+\bb a}^{\ \ -}=0\Ra W^{--}=-\2 \Dp W^{4-}
\end{equation}
\begin{equation}
B_{++--\gd-\bb a}^{\ \ -}=0\Ra \Dp W^{--}+ \Dm W^{++}=0
\end{equation}
While (69) determines $W^{++}$ in terms of $W^{4+}$ and (71) determines $W^{--}$ in terms of $W^{++}$, (70) determines $W^{ 4-}$ in terms of $W^{--}$. One can of course check all these relations by directly substituting their expressions in terms of the $V$'s. To constrain $V^{++}$ we impose
$$F_{\dd-\ga\bb a}^{\ \ ++}=0 $$
\begin{equation}
\Ra D_{\al}^+ V^{++}=0
\end{equation}
i.e. $V^{++}$ is analytic (and real)\footnote{We observe that though $V^{++}$ is analytic, $V^{--}$ is not anti-analytic. Thus the symmetry between plus and minus is now broken, but this is the same thing that happens for example in the 1-form.}. It is now easy to check that the  field function $W^{4+}$ is also real and analytic. This analytic field function is of course the precursor of the unconstrained analytic prepotential that will appear in the 4-form and which should give us a new analytic integration formula. To obtain the higher dimensional components of $A_3$ we put further components of $F$ to zero:
\begin{equation}
F_{\dd-(\ga\bb \al)\ad}^{\ \ -+}=0\Ra A_{\ga\bb\bd\al\ad}^+=-i\4\e_{\bd\ad}\e_{\ga(\al}D_{\bb)}^-V^{++}
\end{equation}
\begin{equation}
F_{\dd+(\ga\bb \al)\ad}^{\ \ -+}=0\Ra A_{\ga\bb\bd\al\ad}^-=-i\4\e_{\bd\ad}\e_{\ga(\al}D_{\bb)}^+V^{--}
\end{equation}
Finally we  have to obtain $A_{cba}$ which we do by imposing the constraint
$F_{\gd-\ga ba}^{\ \ -}=0$. Again after  a  fair share of tedious calculations one obtains
\begin{equation}
A_{\al\ad}=-i\6(2\Dam D_{\al}^+V^{--} + D_{\al}^+\Dap\Dp V^{--} +\Dap D_{\al}^-V^{++})
\end{equation}
The 3-form structure is now complete. One can ideally use Bianchi identities to compute the rest of the non-zero components of the field strength in terms of the field function, but since we will discuss these equations  when we go to 4-form, we leave this  exercise for the moment. 

It is clear from our 3-form analysis that this follows the chain of the forms that we discussed in the earlier section. (That explains the name) The prepotentials match exactly the field functions of the 2-form along with the relations among them. The higher components also look the same. One should ideally now study the multiplet that we have obtained from our 3-form analysis, but since it would be  a  digression for this paper we leave it for future research.
\vspace{5mm}
\\
{\bf The ``New'' 3-form:} If we look at the constraining procedure of the chain-3-form we realise that we used 2 complex gauge fixing conditions to pinpoint a single prepotential. This suggests that there are possibly 2 other independent prepotentials to be discovered. We have indeed obtained at least one of them, which we now present as the new 3-form. As before we start by imposing constraints and use complex gauge fixing to get rid of certain components of $A'_3$:
$$F_{\de\ga\bb\al}^{++++}=0\Ra A_{\ga\bb\al}^{+++}=0,\ \ (\La_{\bb\al}^{++})$$
$$F_{\de\ga\bb\al}^{-+++}=0\Ra A_{\ga\bb\al}^{-++}=0,\ \ (\La_{\bb\al}^{-+})$$
$$F_{\de\ga\bb\al}^{----}=0\Ra A_{\ga\bb\al}^{---}=0,\ \ (\La_{\bb\al}^{--})$$
$$F_{--\ga\bb\al}^{+++}=0\Ra A_{--\bb\al}^{++}=0,\ \ (\La_{--\al}^{+})$$
$$F_{++\ga\bb\al}^{---}=0\Ra A_{++\bb\al}^{--}=0,\ \ (\La_{++\al}^{-})$$
$$F_{++\ga\bb\al}^{+++}=0\Ra A_{++\bb\al}^{++}=0,\ \ (\La_{++\al}^{+})$$
$$A_{++--\de}^{-}=0,\ \ (\La_{--\de}^{-})$$
$$A_{++--\dd-}=0,\ \ (\La_{--\dd+})$$
$$A_{--\bd+\ad+}=0,\ \ (\La_{\bd-\ad+})$$
At this point we impose the reality condition. Then  we impose 
$$F_{++--\bb\al}^{++}=F_{++--\bb\al}^{--}=0\Ra A_{--(\bb\al)}^{-+}=A_{++(\bb\al)}^{-+}=0$$ 
and the prepotentials appear as
\begin{equation}
 A_{--\bb\al}^{-+}\equiv\e_{\bb\al}V^{++}
\end{equation}
\begin{equation}
A_{++\bb\al}^{-+}\equiv\e_{\bb\al}V^{--}
\end{equation}
Further $F_{++--\bb\al}^{-+}=0$  gives us the relation between the two:
\begin{equation}
\Dm V^{++}-\Dp V^{--}=0
\end{equation}
We constrain the prepotential by imposing $$F_{--\ga\bb\al}^{-++}=0$$
\begin{equation}
\Ra D_{\al}^+V^{++}=0
\end{equation}
At this point we can choose to work with real $V^{++} $ which implies\footnote{We could have as well chosen to work with an imaginary $V^{++}$, which will then give an independent solution for the 3-form. However this would be almost exactly identical to the one we are describing modulo factors of $i$ and $-$ signs.}
 that our prepotential $ V^{++}$ is analytic. We now carry on constraining and gauge fixing in  a  straightforward way till we obtain all the non-zero components of $A'_3$, but we will spare the details and enumerate the results: 
\begin{equation}
A_{\ga\bb\al}^{--+}=\e_{(\ga\al}A_{\bb)}^-
\end{equation}
where
\begin{equation}
A_{\bb}^-=-D_{\bb}^+V^{--}
\end{equation}
\begin{equation}
A_{\gd+\bb\al}^{\ \ -+}=\e_{\bb\al}D_{\gd-}V^{--}
\end{equation}
\begin{equation}
A_{\gd-\bd+\al}^{\ \ \ \ -}=\e_{\gd\bd}D_{\al}^+V^{--}
\end{equation}
\begin{equation}
A_{\ga\bb a}^{-+}=\e_{
\bb\al}A_a
\end{equation}
and
\begin{equation}
A_{\al\ad}=i\2 D_{\al}^+\Dam V^{--}
\end{equation}
It is difficult to miss the striking resemblance that this multiplet has with the 1-form, which hints at  a  possible duality between the two. The first nonzero components of field strength appear with four spinor indices: 
\begin{equation}
F_{\de\ga\bb\al}^{--++}\equiv\e_{\de(\al}\e_{\bb)\ga}W
\end{equation}
and
\begin{equation}
F_{\dd-\gd+\bb\al}^{\ \ \ \ \ -+}=\2\e_{\dd\gd}\e_{\bb\al}(W+W^{\x}),
\end{equation}
\begin{equation}
W=D^{+2}V^{--}
\end{equation}
It is not difficult to check that $W$ is chiral and is independent of the harmonic co-ordinates. The argument runs exactly similar to  the 1-form case. This field function, as one can guess, will become the chiral prepotential for the 4-form, and is eventually going to give us the chiral integration formula. We also make  a  curious observation that we have the field function appearing at two different places. The reason is because we worked with  a  real $V^{++}$ but we could have as well  worked with an imaginary  $V^{++}$, the solution being  a  linear combination of the two, and then we should find that two different combinations of real $V^{++}$ and imaginary $V^{++}$ appear in $F_{\de\ga\bb\al}^{--++}$ and $F_{\dd-\gd+\bb\al}^{\ \ \ \ \ -+}$. So essentially we have two independent field functions corresponding to two independent real degrees of freedom in the prepotential.

One can now go on computing the higher field components in terms of the field function(s) using Bianchi identities, but since we will discuss all that in detail when we go to 4-form, we close this chapter.

To summarize, we have seen two (three real) independent prepotentials appearing in the 3-form, which gives rise to different field strengths (which proves for example that they are not gauge equivalent), with different field functions, one being analytic, the other being chiral. However, there is quite  a  bit of resemblance, too, in the prepotentials; both for example are real, analytic and unconstrained with charge 2. Maybe there is something there to explore.

\begin{center} {\bf 5. HIGHER FORMS AND ACTION FORMULAS} \end{center}
{\bf The Ectoplasmic Integration Formula:} The idea of ``ectoplasmic integration formula'' is really quite simple (never mind the name). The component action of  a  supersymmetric theory can be written as an integral over space-time, which implies 
$$ S= \int d^4x (superfields) $$
but superfields are functions of $x,\te$?.
When/how is it that the integral does not depend on $\te$ co-ordinates? This is where the superforms come into play. Consider for example  a  4-form $A_4$ with the field strength $F_5 = dA_4 = 0$ (maximally constrained) and its gauge transformation being given by $\delta_g A_4= d\Lambda_3$.  Then by ``generalized'' Stoke's Theorem the action defined as $S=\int_{bos}A_4$ (integration over the bosonic subspace) will be independent of the rest (in this case $\te$) of the co-ordinates as well. Thus if we can find  a  non-trivial $A_4$, i.e. it is not  a  pure gauge (in which case $S$ will be zero) with zero field strength, then we know how to construct  a  component (bosonic) action in terms of the prepotential of $A_4$. In practice, the components of $A_4$ will be given in terms of a prepotential $V_4$, a  superfield of a definite ``type''. The action formula now reads as
$$S=\int_{bos} A_4(V_4)$$
where $V_4$ is composed from the gauge invariant supersymmetric objects in the theory. An example will perhaps illustrate the matter.

Consider the N=1 case where we indeed have  a  4-form (see N=1 table) with zero field strength. Thus we can write down the action formula (which is also known as the ectoplasmic integration formula) in terms of the 4-form chiral prepotential $V_4$:
 $$
S=\int_{bos}A_4=\int dx^m\wedge dx^n\wedge dx^p\wedge dx^q A_{qpnm}$$
\begin{equation}
= \int dx^m\wedge dx^n\wedge dx^p\wedge dx^q e_m^{\ a} e_n^{\ b} e_p^{\ c} e_q^{\ d} A_{dcba}
\end{equation}
 or
$$ 
S=\int dx^0\wedge dx^1\wedge dx^2\wedge dx^3 \e^{abcd} A_{dcba}\ \  (e_m^{\ a}=\de_m^a) \Ra$$ 
$$( A_{dcba}\equiv\e_{dcba}A)$$
\begin{equation}
S\sim \int d^4x\ A 
\sim \int d^4x(D^2V_4-\D^2\overline{V_4})
\end{equation}
where $V_4$ is a chiral superfield.
 Now we can apply (90) to write down the actions for the various multiplets. For example consider the vector multiplet (1-form): The prepotential is  a  real superfield $V=\overline{V}$ which has  a  chiral spinorial  field function $W_{\al}$ (the gauge invariant supersymmetric object). Then we know $S = S(W^{\al})$ and looking at (90) we realize that all we have to do is to construct  a  chiral scalar ($V_4$) out of  $W_{\al}$. Easily one finds
$$
V_4=W_{\al}W^{\al}$$ and the full action is then given by\footnote{Note $\bar{V_4}=\overline{
W_{\al}W^{\al}}=-\overline{W}_{\al}\overline{W}^{\al}$}
$$ S= \int d^4x [D^2(W_{\al}W^{\al}) + \D^2(\overline{W}_{\al}\overline{W}^{\al})], \ W^{\al}= i\D^2D^{\al}V$$
The expression of the action in terms of the prepotential $V$ is clearly quite  nontrivial, but we have seen using the superform technique how elegantly we arrive at it. In the first step one  obtains the gauge invariant, supersymmetric objects $W_{\al}$ in terms of $V$, and in the next step one finds out how to write down the  action in terms of $W_{\al}$. This procedure has already been discovered for the N=1 case \cite{ecto}. In fact, as is evident,  a  rigid supersymmetry action formula is not difficult to obtain, and this elaborate procedure might seem superfluous, but its real use comes when we look into supergravity, because it is not at all obvious how the vielbein  couples to the multiplet (field functions). However, the ectoplasmic integration formula provides  a  definite natural expansion of the action in terms of the gravitino and gravitini fields \cite{ecto}. In the supergravity case (89) is modified because the vielbein fields are no longer kronecker deltas. Its nontrivial components, namely $e_m^{\ a}\equiv e_m^{\ a}(x,\te)$ and $ \psi_m^{\ \al}=e_m^{\ \al}(x,\te)$ which contain at $\te=0$ the graviton and the gravitino fields, appear in the expansion.
$$
S=\int_{bos}A_4=\int dx^m\wedge dx^n\wedge dx^p\wedge dx^q A_{qpnm}$$
$$= \int dx^m\wedge dx^n\wedge dx^p\wedge dx^q e_m^{\ A} e_n^{\ B} e_p^{\ C} e_q^{\ D} A_{DCBA}$$
$$=\int dx^m\wedge dx^n\wedge dx^p\wedge dx^q (e_m^{\ a} e_n^{\ b} e_p^{\ c} e_q^{\ d} A_{dcba}+\psi_m^{\ \al}e_n^{\ b} e_p^{\ c} e_q^{\ d} A_{dcb\al}$$
$$+\psi_m^{\ \al}\psi_n^{\ \bb} e_p^{\ c} e_q^{\ d} A_{dc\bb\al}+\psi_m^{\ \al}\psi_n^{\ \bb} \psi_p^{\ \ga} e_q^{\ d} A_{d\ga\bb\al}+\psi_m^{\ \al}\psi_n^{\ \bb} \psi_p^{\ \ga} \psi_q^{\ \de} A_{\de\ga\bb\al}$$)
 The components $A_{DCBA}$ can be obtained in terms of the pre-potential $V_4$ by solving the constraint $F=dA=0$, in the now curved background.  It is therefore not surprising that people have used these techniques to obtain supergravity action formulas for the N=1 case \cite{ecto}.   
Our aim is to apply it to N=2 supersymmetry (this paper) and finally to supergravity.
\vspace{5mm}
\\
{\bf Actions in Harmonic Super-Space:} In harmonic superspace, apart from the four space-time co-ordinates, there are two spherical bosonic co-ordinates, and thus there is an arbitrariness as to which bosonic subspace to integrate over while writing down the actions. One can just integrate over the space-time part, in which case the action has to be independent of not only the fermionic but also the spherical co-ordinates. On the other hand one can integrate over the entire bosonic space, or even perform  a  ``contour integral'' in the harmonic part ($S^2$). In the literature all these formulas are known and have been used. While for vector multiplets one usually uses the 4-dimensional chiral integration \cite{harm}, for scalar (and sometimes tensor) multiplets one has to take recourse to the 6-dimensional analytic integration \cite{harm}. The contour integral is possibly the best way of dealing with the tensor multiplets, but this has been studied in  a  slightly different harmonic space, and so the contour integral formula is also known in this different space. However,  a  translation (or mapping) exists between the two spaces \cite{trans},  which will allow us to compare our results with the contour integral formula. 

If the ectoplasmic approach is to work, then we should be able to find non-trivial   4, 5 and 6-forms with zero field strength which will  give us the 4, 5 and 6-dimensional action formulas, respectively. This is precisely what we set out to do, but one can also ask other questions: For example, the formulas that currently exist in literature, are they all? We have seen in our 3-form analysis that there is more than one independent prepotential and this suggests that the same is possibly true for 4,5 and 6-forms. Then we should be able to discover new formulas! Is there  a  relation between these formulas? For example, is it possible to ``dimensionally reduce'' the six dimensional action formula, to say,  a  five dimensional integral? In our analysis we will partially resolve these issues but it is to be emphasized that the higher-form structure in harmonic superspace is quite rich and needs to be explored further. 
\vspace{5mm}
\\
{\bf The 4-Form:} Since our main aim is not to find  multiplets but specifically just to find non-trivial forms with zero field strength, we can take  a  short cut and directly start with an ansatz for the prepotential. Since  we know that we are looking for {\em scalar} prepotentials of {\em dimension 2} \footnote{Since the action is dimensionless and there is an integral over $d^4x$ which contributes to dimension $-4$, the integrand has to have dimension 4. On the other hand we know that whether we are in chiral subspace or analytic subspace, there are half the number of fermionic co-ordinates which have to be eliminated  using covariant spinor derivatives. We have to have 4 of them, which contribute  a  total of dimension 2, which implies the prepotential has to be of dimension 2 as well.}, this approach greatly reduces calculations. It automatically leaves us with    few possible choices where such  a  prepotential can occur, and we start working from there. For example it is easy to check that for the 4-form there are 3 possible choices:\\
(a) $$A_{--\dd-\ga a}^{\ \ +}=\e_{\dd\ad}\e_{\ga\al}V^{4+}$$
$$A_{--\dd+\ga a}^{\ \ +}=\e_{\dd\ad}\e_{\ga\al}V^{++}=-A_{--\dd-\ga a}^{\ \ -}$$
$$A_{--\dd+\ga a}^{\ \ -}=\e_{\dd\ad}\e_{\ga\al}V$$
$$A_{++\dd-\ga a}^{\ \ +}=\e_{\dd\ad}\e_{\ga\al}V'$$
$$A_{++\dd-\ga a}^{\ \ -}=\e_{\dd\ad}\e_{\ga\al}V^{--}=-A_{++\dd+\ga a}^{\ \ +}$$
\begin{equation}
A_{++\dd+\ga a}^{\ \ -}=\e_{\dd\ad}\e_{\ga\al}V^{4-}
\end{equation}
(b)
\begin{equation}
A_{\de\ga\bb\al}^{--++}=\e_{(\de\bb}\e_{\ga)\al}V
\end{equation}
(c)
\begin{equation}
A_{\dd-\gd+\bb\al}^{\ \ \ \ -+}=\e_{\dd\gd}\e_{\bb\al}V
\end{equation}
Now we treat each of these sets as ``to-be'' independent prepotentials (nontrivial) with zero field strength, which will then give us the ectoplasmic integration formulas.  Of course the most general solution will be  a  linear combination of all of these solutions and may be more, but those extra bits are not interesting to us. With these general remarks we embark to solve the 4-form, case by case. 

{\bf (a) The ``chain'' 4-form:} So we start with (91) and assume that all other components of $A_4$ which are of dimension 2 or lower are zero. Let us first investigate whether we still have any residual gauge symmetry, and  proceed to fix them:
$$\de  A_{--\dd+\ga a}^{\ \ -}= \Dp\La_{\dd+\ga a}^{\ \ -} -(\La_{\dd-\ga a}^{\ \ -} -\La_{\dd+\ga a}^{\ \ +})\Ra$$ 
\begin{equation}
\de V=\Dp\La_{++} -(\La+\La^{\x})
\end{equation}
 Similarly we obtain\footnote{One can check that only the antisymmetric combination does not disturb the ansatz.}
\begin{equation}
\de V'=\Dm\La_{--} +(\La+\La^{\x})
\end{equation}
Now if we look at components of the form $A_{\dd\ga\bb a}$ then dimensional analysis gives away  their structure, viz.,
$$
A_{\dd-\ga\bb a}^{\ \ ++}=\e_{\dd\ad}\e_{(\ga\al}V_{\bb)}^{3+}
$$
$$A_{\dd-\ga\bb \al\ad}^{\ \ -+}=\e_{\dd\ad}\e_{\ga\al}V_{\bb}^+ + \e_{\dd\ad}\e_{(\ga\bb}K_{\al)}^+
$$
$$A_{\dd+\ga\bb a}^{\ \ ++}=\e_{\dd\ad}\e_{(\ga\al}V_{\bb)}^{'+}$$
$$A_{\dd-\ga\bb a}^{\ \ --}=\e_{\dd\ad}\e_{(\ga\al}V_{\bb)}^{'-}$$
$$A_{\dd+\ga\bb \al\ad}^{\ \ -+}=\e_{\dd\ad}\e_{\bb\al}V_{\ga}^- + \e_{\dd\ad}\e_{(\bb\ga}K_{\al)}^-
$$
\begin{equation}
A_{\dd+\ga\bb a}^{\ \ --}=\e_{\dd\ad}\e_{(\ga\al}V_{\bb)}^{3-}
\end{equation}
Then
$$F_{\ed-\de\ga\bb a}^{\ \ +++}=0\Ra V_{\al}^{3+}=0,\ \ (\La_{--})$$
We can first  use $(\La+\La^{\x})$ \footnote{Actually $\La,\La^{\x}$ also contribute to $A_{\dd-\gd+\bb\al}^{\ \ \ \ -+}$ but only in the combination $(\La-\La^{\x})$ and therefore our ansatz is not disturbed} to put $V$ to zero and then use ($\La_{++}$) to make $V'$ independent of $u$'s. We strongly believe that this $u$ independent $V'$ gives rise to an independent prepotential (which can be investigated separately) and hence for this solution we will assume it to be zero. We continue to gauge fix
$$
K_{\al}^+=0,\ \ (\la_{\ga ba}^+=\e_{\bd\ad}\e_{(\bb\ga}\La_{\al)}^+)
$$
$$
K_{\al}^-=0,\ \ (\la_{\ga ba}^-=\e_{\bd\ad}\e_{(\bb\ga}\La_{\al)}^-)
$$
Now the constraints
$$F_{\ed-\dd+\ga\bb\al}^{\ \ \ \ --+}=F_{\ed-\dd+\ga\bb\al}^{\ \ \ \ -++}=0$$
\begin{equation}
\Ra V_{\al}^{'+}+V_{\al}^+=V_{\al}^{'-}+V_{\al}^-=0
\end{equation}
Further
$$F_{\ \ \dd-\ga\bb a}^{--\ \ ++}=0\Ra V_{\bb}^+-V_{\bb}^{'+}=0$$
or using (97) we have
\begin{equation}
V_{\bb}^+=V_{\bb}^{'+}=0
\end{equation}
One can check that all the field strength components of the form $F_{\ed\dd\ga\bb\al}$ are zero. The nonzero components of $A_4$ now look like
\begin{center}
 $$A_{--\dd-\ga a}^{\ \ +}=\e_{\dd\ad}\e_{\ga\al}V^{4+}$$
$$A_{--\dd+\ga a}^{\ \ +}=\e_{\dd\ad}\e_{\ga\al}V^{++}=-A_{--\dd-\ga a}^{\ \ -}$$
$$A_{++\dd-\ga a}^{\ \ -}=\e_{\dd\ad}\e_{\ga\al}V^{--}=-A_{++\dd+\ga a}^{\ \ +}$$
\begin{equation}
A_{++\dd+\ga a}^{\ \ -}=\e_{\dd\ad}\e_{\ga\al}V^{4-}
\end{equation}

$$A_{\dd-\ga\bb a}^{\ \ --}=\e_{\dd\ad}\e_{(\ga\al}V_{\bb)}^{'-}$$
$$A_{\dd+\ga\bb a}^{\ \ -+}=\e_{\dd\ad}\e_{\bb\al}V_{\ga}^-
$$
\begin{equation}
A_{\dd+\ga\bb a}^{\ \ --}=\e_{\dd\ad}\e_{(\ga\al}V_{\bb)}^{3-}
\end{equation}
\end{center}
   Of course we have to now re-express this bunch of $V$'s in terms of  a  single prepotential $V^{4+}$ and as usual, the constraints do the job;
$$ F_{++--\gd-\bb a}^{\ \ +}=0  $$
\begin{equation}
\Ra V^{++}=-\2\Dm V^{4+}
\end{equation}
$$ F_{++--\gd+\bb a}^{\ \ -}=0  $$
\begin{equation}
\Ra V^{--}=-\2\Dp V^{4-}
\end{equation}
 and 
$$ F_{++--\gd-\bb a}^{\ \ -}=0  $$
\begin{equation}
\Ra \Dm V^{++}+\Dp V^{--}=0
\end{equation}
$F_{--\dd-\ga\bb a}^{\ \ ++}=0$ makes our prepotential analytic:
\begin{equation}
D_{\al}^+ V^{4+}=0
\end{equation}
At this point one recognizes the similarity of the 4-form with the field strength of the chain-3-form (see for example (69),(70) and (71))  and thus we realize that this form is indeed the continuation of the long chain which began with the 0-form. In chronological order the constraints below  relate other components of  $A$  to $V^{4+}$:
$$F_{--\dd-\ga\bb a}^{\ \ -+}=0$$
\begin{equation}
\Ra A_{--\ga ba}^{+}=i\4\e_{\bd\ad}\e_{(\bb \ga}D_{\al)}^- V^{4+}
\end{equation}
$$F_{--\dd+\ga\bb a}^{\ \ -+}=0 $$
\begin{equation}
\Ra\Dp V_{\al}^-=D_{\al}^-V^{++}
\end{equation}
which determines $V_{\al}^-$ in terms of $V^{4+}$ implicitly, of course. Continuing in this fashion
$$F_{++\dd+\ga\bb a}^{\ \ -+}=0 $$
\begin{equation}
\Ra V_{\al}^{''3-}=\frac{2}{3}(D_{\al}^+V^{4-}+D_{\al}^-V^{--}+\Dm V_{\al}^-)
\end{equation}
and
\begin{equation}
V_{\al}^{3-}=\frac{2}{3}(-\2 D_{\al}^+V^{4-}+D_{\al}^-V^{--}+\Dm V_{\al}^-)
\end{equation}
where 
\begin{equation}
A_{++\ga ba}^{-}=i\4\e_{\bd\ad}\e_{(\bb \ga}V_{\al)}^{''3-}
\end{equation}
$$F_{++--\ga ba}^{+}=0$$
\begin{equation}
\Ra\Dp V_{\al}^{''-}=\Dm D_{\al}^- V^{4+}  
\end{equation}
\begin{equation}
\Ra V_{\al}^{''-}=-2V_{\al}^{-}
\end{equation}
using (106), where
\begin{equation}
A_{++\ga ba}^{+}=i\4\e_{\bd\ad}\e_{(\bb \ga}V_{\al)}^{''-}
\end{equation}
These constraints have thus determined all the components of $A$ up to dimension two and  a  half. However, there are plenty of other constraints of the form $F_{++--\ga ba}^{\pm}=F_{\pm\pm\dd\pm\ga\bb a}^{\ \ \pm\pm}=0$ still to be satisfied by the 4-form. It turns out that by carefully analyzing the Bianchi identities one finds that if the constraint $F_{--\dd+\ga\bb a}^{\ \ --}=0$ is satisfied, then the rest of them are automatically satisfied, too (in other words they are linearly dependent). Since this constraint is the only non-trivial ``check'', so to speak, of our ansatz, let us verify it in  a  little detail:
$$F_{--\dd+\ga\bb a}^{\ \ --}=\Dp A_{\dd+\ga\bb a}^{\ \ --}-D_{(\ga}^-A_{--\dd+\bb) a}^{\ -}+A_{\dd-\ga\bb a}^{\ \ --}-A_{\dd+(\ga\bb) a}^{\ \ -+}=0$$
 if one uses (97) and (100)
implies 
\begin{equation}
\Dp V_{\al}^{3-}=2V_{\al}^-
\end{equation}
Now 
\begin{equation}
(113) \im
\Dp\Dp V_{\al}^{3-}=\Dp 2V_{\al}^- 
\end{equation}
Moreover if one uses  (108) one finds after  a  little playing around 
\begin{equation}
\Dp V_{\al}^{3-}=\3 (-2V_{\al}^-+2\Dm D_{\al}^-V^{++}+2\Dp D_{\al}^-V^{--}-\Dp D_{\al}^+V^{4-})
\end{equation}
Substituting (108) in (115) and further using (106) then one finds 
\begin{equation}
\Dp(2\Dm D_{\al}^-V^{++}+2\Dp D_{\al}^-V^{--}-\Dp D_{\al}^+V^{4-})=8D_{\al}^-V^{++}
\end{equation}
As one can see,  the left hand side of the equation is pretty non-trivial, and one has to do  a  fair bit of manipulation to check that it is exactly equal to the right hand side with the precise coefficient 8 in front of $D_{\al}^-V^{++}$.
Thus we have now $F_{++--\ga ba}^{\pm}=F_{\pm \pm \dd\pm\ga\bb a}^{\ \ \pm\pm}=0$, which of course determines everything in terms of  a  single analytic prepotential, though not always explicitly. Thus it is  a  formidable task to work out the further details, viz. calculating the other components of $A$, and especially the component $A_{dcba}$, which will then give us  a  4-dimensional new analytic integration formula for an analytic unconstrained superfield $V^{4+}$ (and not the conventional 6-dimensional action formula). Though we are still in the process of doing this calculation it is, we believe,  a  matter only of some more (tedious) calculation, and there are no more non-trivial constraints that will come up and spoil the story. In other words we indeed have  a  non-trivial 4-form with zero field strength, precisely what we need to get ectoplasmic integration formulas.

{\bf (b) The ``chiral'' 4-form:} Our beginning ansatz for this form is given by (92), plus that all other components of $A$ with dimension 2 or less are zero, and further that  components of $A$ of the form $A_{\dd\ga\bb a}=0$. Then using constraints one finds $A_{++--\ga ba}=0$, too. As in chain-4-form we compute one by one the field strength components, some of which are automatically zero, the rest give us constraints when we impose them to be zero. For example, 
 $$F_{--\de\ga\bb\al}^{--++}=0\Ra$$
\begin{equation}
\Dp V=0
\end{equation}
In other words $V$ is constrained, i.e. independent of the harmonic co-ordinates. Further 
$$ F_{\eta\de\ga\bb\al}^{---++}=F_{\eta\de\ga\bb\al}^{--+++}=0$$
\begin{equation}
\Ra D_{\al}^-V=D_{\al}^+V=0
\end{equation}
that is, $V$ is chiral. Our next task of course is to determine the higher components of $A$ in terms of the prepotential $V$, and we set about to accomplish this task:
$$ F_{\ed-\de\ga\bb\al}^{\ -+++}=0\Ra A_{\de\ga\bb a}^{+++}=0$$
$$ F_{\ed+\de\ga\bb\al}^{\ ---+}=0\Ra A_{\de\ga\bb a}^{---}=0$$
$$ F_{\ed-\de\ga\bb\al}^{\ --++}=0$$
\begin{equation}
\Ra A_{\de\ga\bb a}^{-++}=-i\4\e_{(\de\bb}\e_{\al)\ga}\Dem V
\end{equation}
$$ F_{\ed+\de\ga\bb\al}^{\ --++}=0$$
\begin{equation}
\Ra A_{\de\ga\bb a}^{--+}=-i\4\e_{(\bb\de}\e_{\al)\ga}\Dep V
\end{equation}
Continuing in this fashion to higher dimensional components we find
$$F_{\ed+\ga\bb\al a}^{\ \ --+}=0 $$ 
\begin{equation}
\Ra A_{\de\ga ba}^{--}=-\6\e_{\bd\ad}\e_{(\bb\de}\e_{\al)\ga}\D_+^2V
\end{equation}and similarly
$$F_{\ed-\ga\bb\al a}^{\ \ -++}=0 $$ 
\begin{equation}
\Ra A_{\de\ga ba}^{++}=-\6\e_{\bd\ad}\e_{(\bb\de}\e_{\al)\ga}\D_-^2V
\end{equation}
$$F_{\ed-\ga\bb\al a}^{\ \ --+}=0 $$ 
\begin{equation}
\Ra A_{\de\ga ba}^{-+}=\6(\e_{\bd\ad}\e_{(\bb\de}\e_{\al)\ga}\D_-\D_+V+\e_{\bb\al}\e_{\de\ga}\overline{D}_{(\bd-}\overline{D}_{\ad+)}V)
\end{equation}
Next $$F_{\ed+\dd+\ga\bb a}^{\ \ \ \ -+}=0\Ra A_{\dd+\ga ba}^{\ \ -}=0$$ and similarly $$F_{\ed-\dd-\ga\bb a}^{\ \ \ \ -+}=0\Ra A_{\dd-\ga ba}^{\ \ +}=0$$ 
The constraints $F_{\ed-\dd+\ga\bb a}^{\ \ \ \ -+}=F_{\ \ \dd+\ga ba}^{++\ \ -}=0$ together with the $u$-independent part of the gauge form $\La_{cba}$ enables us to put $$A_{\dd+\ga ba}^{\ \ +}=A_{\dd-\ga ba}^{\ \ -}= 0$$
Thus up to the dimension 3 all components of $A$ are determined.
Carrying on
$$F_{\ed+\de\ga ba}^{\ \ -+}=0$$
\begin{equation}
\Ra A_{\de \al\ad}^-=-\4\e_{\de\al}\Dam \D_+^2V
\end{equation}
where
\begin{equation}
A_{\de cba}^-=\e_{fcba}A_{\de}^{-f}
\end{equation}
and similarly
$$F_{\ed+\de\ga ba}^{\ \ -+}=0$$
\begin{equation}
\Ra A_{\de \al\ad}^+=-\4\e_{\de\al}\Dam \D_-^2V
\end{equation}
where
\begin{equation}
A_{\de cba}^+=\e_{fcba}A_{\de}^{+f}
\end{equation}
Finally then one can compute $A_{dcba}$ by looking at the constraint $F_{\dd-\de cba}^{\ \ -}$, for example. One finds after straightforward manipulations
\begin{equation}
A=\6 i(\D_+^2\D_-^2V-D^{-2}D^{+2}V^{\x})
\end{equation}
where $A$ is defined as in (90). One can check that the expression of course is real, as it should be.
We checked that all other field strength components of the chiral  form are indeed   zero, and by construction it is not  a pure gauge.
The integration formula that follows from this form is then (after incorporating the $i$ by redefining $V$ and dropping the irrelevent overall factor)
\begin{equation}
S=\int d^4x \ A=\int d^4x (\D_+^2\D_-^2V+D^{-2}D^{+2}V^{\x})
\end{equation}
where $V$ has to be  a  constrained (117) chiral (118) object.
This is the well known chiral integration formula used for constructing actions from chiral gauge invariant objects, and we have reproduced it using ectoplasmic methods from  a  4-form. For example one can construct the Yang-Mill's action from the chiral field function $W$ as
$$S=\int d^4x (\D_+^2\D_-^2W^2+D^{-2}D^{+2}W^{2\x})$$which in this special case reduces to (31).
Also one can see that this 4-form has the same structure as the field strength of the new-3-form (in fact ``half'' of that), and thus can be considered as  a  continuation of that chain. 

{\bf (c) The ``Real'' 4-form:} The structure of the field function of the new-3-form suggests that the prepotential of the real-4-form (93) will again be  a  chiral and constrained $V$:
\begin{equation}
A_{\dd-\gd+\bb\al}^{\ \ \ \ -+}=\e_{\dd\gd}\e_{\bb\al}(V+V^{\x})
\end{equation}
Since it is unlikely that two different integration formulas will exist for chiral superfields we believe that if one works this one out then we will ultimately obtain the same chiral integration formula (129) that we got from the chiral-4-form. Thus we do not investigate this further, but for completeness sake it might be  a  good future exercise.

Thus we have worked out the 4-form ``almost'' completely, obtained the chiral integration formula and also seen how the structure of the forms propagate as  a  chain starting from the 0-forms. It may be  a  good idea to write down  a  table like for N=1 case,  for the two chains in N=2 that we have seen:

\begin{equation} \begin{array}{lll}
p=0:& A=V& W^{++}=\Dp V\\
p=1: & A_{--}=V^{++},\ A_{++}\equiv V^{--}&W=-\2 D^{+2}V^{--}\\
 &[D_{--}V^{--}=D_{++}V^{++}]&\\
 &A_{\al}^-=D_{\al}^+A_{++},\ A_{\al\bd}=-i\2 D_{\al}^+\D_{\bd -}A_{++}& \\
p=2:&  A_{\bb\al}^{-+}=\e_{\bb\al}V, \ A_{\bb\al\ad}^+=i\2 \e_{\bb\al}D_{\ad -}V& W^{++}=-i\frac{1}{4}(\D_{-}^2V\\
 & A_{\bb\bd\al\ad}=\frac{1}{8}(\e_{\bb\al}\D_{(\bd -}\D_{\ad) +}V+s.c.) & \ \ 
 \ +D^{+2}V^{\x})\\
p=3:&  A_{\gd - \bb\al\ad}^{\ +}=\e_{\gd\bd}\e_{\bb\al}V^{++},\ A_{\gd +\bb\al\ad}^{\ -}= -\e_{\gd\bd}\e_{\bb\al}V^{--}& W^{4+}=D^{++}V^{++}\\
 & A_{\gd - \bb,\al\ad}^{\ -}=-A_{\gd +, \bb,\al\ad}^{\ +}=\e_{\gd\bd}\e_{\bb\al}V& \\
 &[\Dm V^{++}=-\Dp V^{--}=2V]&\\  
 &A_{\ga\bb\bd\al\ad}^{\pm}=\mp i\frac{1}{4}\e_{\bd\ad}\e_{\ga(\al}D_{\bb)}^\mp V^{\pm\pm}& \\
 &  A_{\ga\ad}=-\frac{i}{16}(D_{\ga}^-\D_{\ad +}V^{++}-D_{\ga}^-\D_{\ad -}V^{--})&\\
p=4:&  A^{\ \pm}_{\mp\mp\gd \mp\bb\al\ad}=\e_{\gd\bd}\e_{\bb\al}V^{4\pm} & W=0\\
& A^{\ \mp}_{\mp\mp\gd \mp\bb\al\ad}=\2\e_{\gd\bd}\e_{\bb\al}D^{\pm\pm}V^{4\mp} &\\
 & A_{\dd+\ga\bb a}^{\ \ -+}=\e_{\dd\ad}\e_{(\ga\al}V_{\bb)}^{-},\ [\Dp V_{\al}^-=D_{\al}^-V^{++}]&\\
 &A_{\dd-\ga\bb a}^{\ \ -+}=-\e_{\dd\ad}\e_{(\ga\al}V_{\bb)}^{-},\ A_{\dd+\ga\bb a}^{\ \ --}=\e_{\dd\ad}\e_{(\ga\al}V_{\bb)}^{3-}&\\
& [V_{\al}^{3-}=\frac{2}{3}(-\2 D_{\al}^+V^{4-}+D_{\al}^-V^{--}+\Dm V_{\al}^-)]&\\
 & A^{+}_{--\ga ba}=-\frac{i}{4}\e_{\bd\ad}\e_{\ga(\bb}D_{\al)}^-V^{4+}&\\
 & A^{\ -}_{++\ga,\bb\bd,\al\ad}=\frac{i}{4}\e_{\bd\ad}\e_{\ga(\bb}V_{\al)}^{''3-}&\\
 &[V_{\al}^{''3-}=\frac{2}{3}(D_{\al}^+V^{4-}+D_{\al}^-V^{--}+\Dm V_{\al}^-)]&  
\end{array}
\end{equation}

In this table, except for the 2-form prepotential, which is chiral and constrained, all the other prepotentials are unconstrained analytic and real. 

The second table is quite short since it starts at the 3-form: 
\begin{equation}
\begin{array}{lll}
p=3:&A_{--\bb\al}^{-+}=\e_{\bb\al}V^{++},\ A_{++\bb\al}^{-+}=\e_{\bb\al}V^{--}& W=D^{+2}V^{--}\\
 &[D^{++}V^{--}=D^{--}V^{++}],\ A_{\ga\bb\al}^{--+}=-\e_{(\ga\al}D_{\bb)}^+V^{--}&\\
 &A_{\gd+\bb\al}^{--+}=-\e_{\bb\al}\Dgm V^{--},\ A_{\ga\bb a}^{-+}=\e_{\ga\bb}V_a& \\
 & [V_{\al\ad}=i\2 D_{\al}^+\Dam V^{--}]&\\
p=4: & A_{\de\ga\bb\al}^{--++}=\e_{(\de\bb}\e_{\ga)\al}V,\ A_{\de\ga\bb a}^{-\pm+}=-i\4\e_{(\bb\de}\e_{\al)\ga}\overline{D}_{\ed\mp} V& W=0\\
 & A_{\de\ga ba}^{\pm\pm}=-\6\e_{\bd\ad}\e_{(\bb\de}\e_{\al)\ga}\D_{\mp}^2V&\\
 & A_{\de\ga ba}^{-+}=\6(\e_{\bd\ad}\e_{(\bb\de}\e_{\al)\ga}\D_-\D_+V+\e_{\bb\al}\e_{\de\ga}\overline{D}_{(\bd-}\overline{D}_{\ad+)}V)&\\
 & A_{\de cba}^{\pm}=\e_{fcba}A_{\de}^{\pm f},\ [A_{\de\al\ad}^{\pm}=-\4\e_{\de\al}\overline{D}_{\ad\pm} \D_{\mp}^2V]&\\
 & A_{dcba} =i\6\e_{dcba}(\D_+^2\D_-^2V-D^{-2}D^{+2}V^{\x})&
\end{array}
\end{equation}
{\bf The 5-Form:} Our sole aim for the 5 and 6-forms is to obtain nontrivial $A$'s with zero field strength to obtain the integration formulas, and in particular the already known action formulas as mentioned in the introduction. The complete 5 and  6-form structure we believe is pretty rich, and it will take quite some effort to uncover all that they have to offer. In this paper we only find  a  special solution (for each case) which as we shall see will reproduce the known ones. For the 5-form we start with the following ansatz for the prepotential:
$$A_{\mp\mp\dd\mp\ga\bb a}^{\mp\mp}=\e_{\dd\ad}\e_{(\ga\al}V_{\bb)}^{'\pm}$$
$$A_{\mp\mp\dd\pm\ga\bb a}^{-+}=\e_{\dd\ad}\e_{(\ga\al}V_{\bb)}^{\pm}$$
\begin{equation}
A_{\mp\mp\dd\pm\ga\bb a}^{\mp\mp}=\e_{\dd\ad}\e_{(\ga\al}V_{\bb)}^{''\mp}
\end{equation}
As usual we keep computing the field strength components and check whether they are zero or not. If they already are, then we do not get any constraints, but whenever we have  a  non-zero expression, the condition that it has to vanish gives us  a  relation/constraint. Sparing the unimportant details here we only mention the non-trivial constraints. 
$$F_{\mp\mp\ed-\dd+\ga\bb\al}^{\ \ \  \ -\mp +}=0$$
\begin{equation}
\Ra V_{\al}^{\pm}+V_{\al}^{'\pm}=0
\end{equation}
$$F_{++--\dd\mp\ga\bb a}^{\ \ \mp\mp}=0$$
\begin{equation}
\Ra -D^{\mp\mp}V_{\al}^{\pm} +2V_{\al}^{\mp} +V_{\al}^{''\mp}=0
\end{equation}
$$F_{++--\dd\mp\ga\bb a}^{\ \ \pm\pm}=0$$
\begin{equation}
\Ra D^{\pm\pm}V_{\al}^{''\pm}=0
\end{equation}
These are the only independent constraints that one obtains, and clearly there is some freedom in the solution still left (probably because we did not bother to fix the gauge). We here make  a  simple ansatz:\\
\begin{equation}
V_{\al}^{\pm}=k^{\pm}V_{\al}^{''\pm}
\end{equation}
It turns out that our ansatz solves the constraints (134), (135) and (136) provided 
\begin{equation}
3k^+k^- +2(k^+ +k^-) +1=0
\end{equation}
We now look at some more constraints:
$$F_{--\ed+\de\ga\bb a}^{\ \ ---}=0\Ra D_{(\bb}^-V_{\al)}^{''-}=0$$
\begin{equation}
\Ra V_{\al}^{''-}=D_{\al}^-V
\end{equation}
Similarly we get, 
$$F_{++\ed-\de\ga\bb a}^{\ \ +++}=0\Ra D_{(\bb}^-V_{\al)}^{''+}=0$$
\begin{equation}
\Ra V_{\al}^{''+}=D_{\al}^+V'
\end{equation}
Consider
$$F_{--\ed+\de\ga\bb a}^{\ \ --+}=0 $$
\begin{equation}
\Ra 2i A_{--\de\ga ba}^{--}=-\e_{\bd\ad}(\e_{(\ga\al}D_{\de)}^-V_{\bb}^++ \e_{\bb\al}D_{(\de}^-V_{\ga)}^+ +\e_{(\ga\al}D_{\bb}^+V_{\de)}^{''-})
\end{equation}
From symmetry properties  we see the $[\bb\al]$ part of $A_{\ \ \de\ga ba}^{++--}$ is zero 
\begin{equation}
 3D_{(\bb}^-V_{\al)}^+ +D_{(\bb}^+V_{\al)}^{''-}=0
\end{equation}
Using (135) (137) and (142) one then can prove
\begin{equation}
D_{(\bb}^+V_{\al)}^{''-}=0\Ra V_{\al}^{''-}=D_{\al}^+V^{--}
\end{equation}
However we also know (136), i.e. $V_{\al}^-$ contains only one $u$ in its harmonic expansion, and thus we can always find  a  $V^{--}$ such that (143) is satisfied while
\begin{equation}
\Dm V^{--}=0
\end{equation}
but then hitting (144) with $\Dm$ one finds
\begin{equation}
D_{\al}^- V^{--}=0
\end{equation}
Using similar arguments $$F_{--\nd+\de\ga\bb a}^{\ \ --+}=0 $$ 
\begin{equation}
\Ra V_{\al}^{''+}=D_{\al}^-V^{++}
\end{equation}
where
\begin{equation}
D_{\al}^+ V^{++}=\Dp V^{++}=0
\end{equation}
Finally we have obtained our ``true'' prepotential $V^{++}$. Of course $V^{++}$ and $V^{--}$ are completely determinable in terms of each other. (In fact one can even obtain expressions for $V$ and $V'$.)

Let us now return back to (141) to evaluate $A_{--\de\ga ba}^{--}$, which in a straightforward way yields
\begin{equation}
A_{--\de\ga ba}^{--}=-i\4\e_{\bd\ad}\e_{(\ga\al}\e_{\de)\bb}(k^+ D^{-2}V^{++}-D^{+2}V^{--})
\end{equation}
and similarly one obtains 
\begin{equation}
A_{++\de\ga ba}^{++}=-i\4\e_{\bd\ad}\e_{(\ga\al}\e_{\de)\bb}(k^- D^{+2}V^{--}-D^{-2}V^{++})
\end{equation}
while $$F_{--\ed-\de\ga\bb a}^{\ \ --+}=0\Ra A_{++\de\ga ba}^{-+}=0$$ 
Continuing we further discover
$$F_{--\ed-\dd-\ga\bb a}^{\ \ \ \ -+}=0\Ra A_{--\dd-\ga ba}^{\ \ +}=0$$
 $$F_{++\ed+\dd+\ga\bb a}^{\ \ \ \ -+}=0\Ra A_{++\dd+\ga ba}^{\ \ -}=0$$
$$F_{--\ed+\dd+\ga\bb a}^{\ \ \ \ -+}=0$$
\begin{equation}
\Ra A_{--\dd+\ga\bb\al}^{\ \ -}=-i\4\e_{\ga(\al}(D_{\bb)}^+\Dem V^{--}-k^+\Dep D_{\bb)}^-V^{++})
\end{equation}
where 
\begin{equation}
A_{--\dd+\ga ba}^{\ \ -}=\e_{\bd\ad}A_{--\dd+\ga\bb\al}^{\ \ -}+\e_{\bb\al}A_{--\gd+\eta\bb\al}^{\ \ -\x}
\end{equation}
A similar calculation performed with $F_{++\ed-\dd-\ga\bb a}^{\ \ \ \ -+}=0$ yields
\begin{equation}
A_{++\dd-\ga\bb\al}^{\ \ +}=-i\4\e_{\ga(\al}(D_{\bb)}^-\Dep V^{++}-k^-\Dem D_{\bb)}^+V^{--})
\end{equation}
where 
\begin{equation}
A_{++\dd-\ga ba}^{\ \ +}=\e_{\bd\ad}A_{++\dd-\ga\bb\al}^{\ \ +}+\e_{\bb\al}A_{++\gd-\eta\bb\al}^{\ \ +\x}
\end{equation}
One can continue to work in this general framework till he reaches $A_{\mp\mp dcba}$ but a clever trick simplifies the calculations greatly: First we realize that using general dimensional arguments  and gauge freedom we can write down $A_{--\dd\pm\ga ba}^{\ \ \pm}$ as
\begin{equation}
A_{--\dd-\ga ba}^{\ \ -}=\e_{\bd\ad}\e_{(\bb\ga}A_{--\al)\ed}+\e_{\bb\al}\e_{(\bd\ed}A_{--\ga\ad)}
\end{equation}
with $A_{--\ga\ad}^{\x}=A_{--\gd\al}$. Then 
$$F_{--\ed-\dd+\ga\bb a}^{\ \ \ \ -+}=0$$
\begin{equation}
\Ra A_{--\ad\al}=i\4 k^+\Dam D_{\al}^- V^{++}
\end{equation}
but now we can choose $k^+=0 \Ra k^-=-\2$ (138). In this ``gauge'' the ``$--$'' sector suddenly becomes very simple. The only dimension 3 components are now given by 
\begin{equation}
A_{--\de\ga ba}^{--}=i\4\e_{\bd\ad}\e_{(\ga\al}\e_{\de)\bb}D^{+2}V^{--}
\end{equation}
\begin{equation}
A_{--\dd+\ga\bb\al}^{\ \ -}=-i\4\e_{\ga(\al}D_{\bb)}^+\Ddm V^{--}
\end{equation}
Now consider the constraint $F_{--\ed-\de\ga ba}^{\ \ --}=0$. Only two terms survive because of the clever choice of the gauge:
$$ F_{--\ed-\de\ga ba}^{\ \ --}=-\Dem A_{--\de\ga ba}^{--}+2iA_{--\de\ga\ed ba}^{-}=0$$
\begin{equation}
\Ra A_{--\de\al\ad}^{-}\sim\e_{\de\al}\Dam D^{+2}V^{--}
\end{equation}
where
\begin{equation}
A_{--\de cba}^{\pm}=\e_{fcba}A_{--\de}^{\pm f}
\end{equation}
Also we have
$$F_{--\ed+\de\ga ba}^{\ \ ++}=0\Ra A_{--\de\al\ad}^{+}=0$$

Finally we consider the constraint $F_{--\ed-\de cba}^{\ \ -}=0$, which also becomes extremely simple and we obtain $A_{--dcba}$:
\begin{equation}
A_{--}\sim \D_-^2D^{+2}V^{--}
\end{equation}
where $A$ is defined as in (90). Thus we can now obtain the integration formula over the ``$--$'' five-dimensional surface: 
\begin{equation}
S=\int \om^{--} d^4x \ A_{--}=\int \om^{--} d^4x \ \D_-^2D^{+2}V^{--}
\end{equation}
where $V^{--}$ is an anti-analytic constrained prepotential.
Though the  ``++'' sector  of the form is now very complicated (because of our choice of gauge) we can actually use another   trick to compute $A_{++dcba}$. Consider the constraint
\begin{equation}
F_{++--dcba}=\Dp A_{++dcba}-\Dm A_{--dcba}=0
\end{equation}
This uniquely determines $A_{++dcba}$ in terms of $A_{--dcba}$, and by inspection we can actually find the solution  modulo the co-efficient
\begin{equation}
A'\sim \D_+^2D^{-2}V^{++}
\end{equation}
which now gives the ``++'' 5-dimensional integration formula:
\begin{equation}
S'=\int \om^{++} d^4x \ A_{++}=\int \om^{++} d^4x \ \D_+^2D^{-2}V^{++}
\end{equation}
where $V^{++}$ is the constrained analytic prepotential. These two formulas are new in this form. As we had mentioned earlier this one-dimensional integration (contour) in the spherical part is almost never used in the usual harmonic space  approach but rather used in the complex CP(1) harmonic space approach. These two spaces however are isomorphic and an elaborate mapping procedure of functions, operators,  etc., between the two spaces is discussed in \cite{trans}. Following the prescription given in the aforementioned paper we could re-write the action formulas (161) and (162) so that we generated the popular contour action formulas. (For details see appendix C.) 
\vspace{5mm}
\\
{\bf The 6-Form:} We start with an ansatz which is very similar to the 5-form ansatz:
\begin{equation}
A_{++--\dd\mp\ga\bb a}^{\pm\pm}=\e_{\dd\ad}\e_{(\ga\al}V_{\bb)}^{3\pm}
\end{equation}
We assume all other components of $A_6\leq \frac{5}{2}$ are zero.  Then
$$F_{++--\ed\pm\de\ga\bb a}^{\pm\pm\pm}=0$$
\begin{equation}
\Ra D_{(\bb}^{\pm}V_{\al)}^{3\pm}=0\Ra V_{\al}^{3\pm}=D_{\al}^{\pm}V^{\pm\pm}
\end{equation}
From now on we will suppress the ``$--++$'' indices on the top since any component which does not contain both the harmonic indices is zero by our ansatz.  We now embark upon expressing higher components using the constraints:
 $$ F_{\ed-\de\ga\bb a}^{\ \ --+}=0 \Ra A_{\de\ga ba}^{-+}=0$$
$$ F_{\ed-\de\ga\bb a}^{\ \ -++}=0  $$
\begin{equation}
\Ra A_{\de\ga ba}^{++}=-i\2\e_{\bd\ad}\e_{(\ga\al}D_{\bb}^-D_{\de)}^+V^{++}
\end{equation}
From symmetry arguments again we see that the $[\bb\al]$ part of $A_{\de\ga ba}^{++}$ is zero. Therefore we get
\begin{equation}
D_{(\bb}^-D_{\ga)}^+V^{++}=0 \Ra D_{\al}^+V^{++}=D_{\al}^-V^{4+}
\end{equation} with
\begin{equation}
D_{(\bb}^-D_{\ga)}^+V^{++}=0 
\end{equation}
We now focus into  a special case which solves (168) and (169), viz.
\begin{equation}
V^{++}=\Dm V^{4+}
\end{equation}
with the prepotential $V^{4+}$ satisfying the conditions
\begin{equation}
D_{\al}^+ V^{4+}=0
\end{equation}
Now
\begin{equation}
A_{\de\ga ba}^{++}=-i\4\e_{\bd\ad}\e_{(\ga(\al}D_{\bb)}^-D_{\de))}^+V^{++}
\end{equation}
By similar reasoning we find
$$ F_{\ed+\de\ga\bb a}^{\ \ --+}=0  $$
\begin{equation}
\Ra V^{--}=\Dp V^{4-}
\end{equation}
\begin{equation}
D_{\al}^- V^{4-}=0
\end{equation}
and
\begin{equation}
A_{\de\ga ba}^{--}=-i\4\e_{\bd\ad}\e_{(\ga(\al}D_{\bb)}^+D_{\de))}^-V^{--}
\end{equation}
Using $F_{\ed-\dd+\ga\bb a}^{\ \ \ \ -+}=0$ and gauge choice $\La_{cba}$ we see
$$A_{\ed\pm\ga ba}^{\ \ \pm}=0$$
Proceeding further on we find
$$F_{\ed\mp\dd\mp\ga\bb a}^{\ \ \ \ -+}=0  $$
\begin{equation}
\Ra A_{\ed\pm\ga ba}^{\ \ \mp}=-i\4(\e_{\bd\ad}\e_{\ga(\al}D_{\bb)}^{\pm}\Dep V^{\mp\mp}+\e_{\bb\al}\e_{\ed(\ad}D_{\ga}^{\pm}\D_{\bd)\pm} V^{\mp\mp})
\end{equation}
Then
$$F_{\ed\mp\de\ga ba}^{\ \ -+}=0$$
\begin{equation}
\Ra A_{\de\al\ad}^{\pm}=i\6\e_{\al\de}D^{\mp 2}\D_{\ad\mp} V^{\pm\pm}
\end{equation}
Finally to obtain $A_{++--dcba}$ we look at 
$$F_{\dd-\de cba}^{\ \ -}=0$$
\begin{equation}
\Ra A\sim (D^{+2}\D_-^2 V^{4-}+D^{-2}\D_+^2 V^{4+})
\end{equation}
where $A$ is as usual given as in (90). Note that here we actually have two independent prepotentials ($V^{4+},V^{4-}$) and hence we actually get two different action formulas:
\begin{equation}
S=\int d^2u d^4x A=\int d^2u d^4x D^{+2}\D_-^2 V^{4-}
\end{equation}
and 
\begin{equation}
S=\int d^2u d^4x A=\int d^2u d^4x D^{-2}\D_+^2 V^{4+}
\end{equation}
One immediately recognizes that these correspond to the usual action formulas for analytic and anti-analytic superfields (for example, see (19)). 

\begin{center}  {\large {\bf SUMMARY AND FUTURE RESEARCH}} \end{center}

We have seen that harmonic superforms like the N=1 superforms describe most of the important multiplets in the N=2 theory, viz. the scalar, vector and tensor multiplets. Proceeding in  a  systematic manner starting from the vielbein fields, we have constructed these supermultiplets (which are described by the prepotential, residing in the forms) and the gauge invariant supersymmetric objects (field-functions) in terms of which we can write down their action. Moreover we have seen that like the N=1 case here also the forms form  a  chain where the field strength of one resembles the prepotential of the next and so on and so forth. However, we also realized that the superform structure is much richer in harmonic superforms, with the possibility of more than one independent prepotential existing at  a  single level, specifically in the higher forms (3 or more). The ``branching off of the chain'' is not completely clear because we believe that the ``tree'' is not complete and more forms are there to be discovered. However, with the forms that we investigated we saw that we could apply the ectoplasmic  ideas to obtain the already known action formulas (one chiral and two analytic)  for the harmonic superspace. Specifically the 4, 5 and  6-forms gave us 4, 5 and 6-dimensional action formulas. Moreover we have also seen the existence of  a  new 4-form.

These results, though encouraging, are  not an end in itself. Firstly one should investigate the tree of forms further and complete it. We believe there are one more 3 and 4-form left, while the 5 and the 6-form are still wide open. If we get all these formulas, we might be able to see certain connections between the two (spherical and contour integral), which among other things should lead to a better understanding of the various dualities  that exist. 

Finally our aim of course will be to carry over all these to curved superspace
 or supergravity. Now that we know where exactly the relevant prepotentials sit in the 4, 5 and 6-forms, it might not be too difficult to generalize the ectoplasmic ideas to supergravity, which certainly will be an important accomplishment.

\begin{center}  {\large {\bf APPENDIX A The star conjugate operation}} \end{center}
   To understand how the star conjugate operator ``$\x$'' works, we have to first understand how the usual complex conjugation works on fermionic objects. We first define:
$$
\overline{(\te^{\al}_i)}=\overline{\te}^{\ad i}
$$
\begin{equation}
\overline{(\overline{\te}^{\ad i})}=\te^{\al}_i
\end{equation}
Then complex conjugation on any arbitrary function is defined by hermitian conjugation. 
\begin{equation}
\overline{(\chi_1\chi_2...\chi_n)}=\overline{\chi_n}...\overline{\chi_2}\ \overline{\chi_1}
\end{equation}
where $\chi_i$ can be either commuting or anticommuting. Next we move on to define complex conjugation of operators (differential) in the following sense: Say, $O$ is a differential operator, then we define its conjugate $\overline{O}$ as
\begin{equation}
\overline{Of}=\overline{O}\ \overline{f}
\end{equation}
where $f$ is a superfield. It turns out that $\overline{O}$ can be different depending upon whether $f$ is commuting or anticommuting. For example, using the defining relation (183) one can check 
$$
\overline{\frac{\p}{\p\te}}=\left\{ \begin{array}{c} -\frac{\p}{\p\tb}\mathrm{\ acting\ on\ bosons}\\
\ \ \frac{\p}{\p\tb} \mathrm{\ acting\ on\ fermions} \end{array}\right.$$
Similar calculations yield (7). 

One can  extend the conjugation operation to superforms as well. We require
 \begin{equation}
\overline{(d(\Lambda))}=d(\overline{\Lambda})
\end{equation}
One can check that a consistent and natural set of rules for conjugation is given by 
$$(\Omega^M f_M)^{\x}=(-)^m \Omega^{M\x}f_M^{\x}$$
\begin{equation}
(\Omega^M\wedge\Lambda^N)^{\x}=(-)^{mn}\Omega^{M\x}\wedge\Lambda^{N\x}
\end{equation}
where $m,n$ takes the value plus or minus 1 depending on whether $M,N$ are bosonic or fermionic respectively. The ``diamond'' operation as defined earlier acts only on the harmonic functions ($u_{\pm}^i \ra \pm u_{\mp}^i$). Hence, the rules  remain the same when we replace the conjugation with star conjugation. Equipped with these rules, it is straightforward to calculate how the star conjugation acts on specific differential forms and operators.

\begin{center} {\large {\bf APPENDIX B The contour integral from 5-form harmonic superspace integral}} \end{center}
  We obtained the 5-dimensional action formula as
\begin{equation}
S=\int \om^{++} d^4x \ A_{++}=\int \om^{++} d^4x \ \D_+^2D^{-2}V^{++}
\end{equation}
However, in the literature one is more familiar with the contour integral formula \cite{cp11} for writing down actions in CP(1) harmonic spaces. Fortunately a translation between the two spaces exists \cite{trans}, which we use to obtain the familiar contour integral action starting from (186). In the northern patch of the sphere (coset space) the translation is given by 
\begin{equation}
u_i^{\ u}=\frac{1}{\sqrt{1+t\bar{t}}}\left( \begin{array}{cc} t&1\\
                                               -1 &\bar{t}
\end{array} \right), \ u_u^{\ i}=\frac{1}{\sqrt{1+t\bar{t}}}\left( \begin{array}{cc} \bar{t} & -1\\
                                                                    1  & t
\end{array} \right),
\end{equation}
where $t$ and $\bar{t}$ can be viewed as co-ordinates describing the coset manifold. With this translation it is easy to compute $\om^{++}=u_-^idu_i^+$, and we obtain
\begin{equation}
\om^{++}=\frac{dt}{1+t\bar{t}}
\end{equation}
It was shown in \cite{trans} that
\begin{equation}
\D_+^2D^{-2}V^q=(\bar{u}^{+1})^4 \frac{(1+t\bar{t})^4}{t^2}D^4 V^q
\end{equation}
where corresponding to any $V^q(y^{\dot{m}})$ one can define a $V^{(q)}_N(t,\bar{t})$ \cite{trans}:
\begin{equation}
V^q(y^{\dot{m}})=(u^{+1})^q V^{(q)}_N(t,\bar{t})
\end{equation}
Further we have 
\begin{equation}
\Dp V^q(y^{\dot{m}})=0\Ra \frac{\p}{\p\bar{t}}V^{(q)}_N(t,\bar{t})=0
\end{equation}
Thus, we finally have, substituting (188), (189) and (190) into (186)\footnote{It is intuitively clear why the integral over $dt$ has to be a contour integral, because it is spanning a compact manifold, but perhaps a more rigorous mathematical understanding would be nice.} 
$$
\int \om^{++} d^4x \ \D_+^2D^{-2}V^{++}=\int d^4x\oint (\frac{dt}{1+t\bar{t}})((\bar{u}^{+1})^4 \frac{(1+t\bar{t})^4}{t^2}D^4)(u^{+1})^2 V^{(q)}_N(t)
$$
Substituting $u^{+1}=-u_-^1$  one then finds
\begin{equation}
S=\int \om^{++} d^4x \ \D_+^2D^{-2}V^{++}=\int d^4x\oint\frac{dt}{t^2}D^4V^{(q)}_N(t)\equiv \int d^4x\oint\frac{dt}{t}D^4L(t)
\end{equation}
the familiar contour integral action formula.


\begin{thebibliography}{99}

\bibitem{nogo} M. Ro\v{c}ek, W. Siegel, {\it Phys.Lett.} {\bf 105 B} (1981) 275
\bibitem{harm} A.Galperin, E.Ivanov, S.Kalitzin, V.Ogievetsky, E.Sokatchev, {\it Class. Quantum Grav.} {\bf 1} (1984) 469;
A. Galperin, E. Ivanov, S. Kalitzin, V. Ogievetsky, E. Sokatchev, {\it Class. Quantum Grav.} {\bf 2} (1985) 155
\bibitem{ecto} S.J. Gates, Jr., hep-th/9709104, Ectoplasm has no topology: the prelude, in Supersymmetries and quantum symmetries, proc., July 22-26, 1997 Dubna ({\it Lecture notes in
physics}, {\bf v. 524}), eds. J. Wess and E.A. Ivanov (Springer, 1999) p. 46;
S.J. Gates, Jr., M.T. Grisaru, M.E. Knutt-Wehlau, and W. Siegel, hep-
th/9711151, {\it Phys. Lett.} {\bf 421} (1998) 203; S.J. Gates, Jr., {\it Nucl.Phys.} {\bf B 541} (1999) 615
\bibitem{cp11} A. Karlhede, U. Lindstr\"{o}m, M. Ro\v{c}ek, {\it Phys.Lett.} {\bf 147 B} (1984) 297; F. Gonzalez-Rey, U. Lindstr\"{o}m, M. Ro\v{c}ek, R. von Unge, S. Wiles, {\it Nucl. Phys.} {\bf B 516} (1998) 426; F. Gonzalez-Rey,  R. von Unge,  {\it Nucl. Phys.} {\bf B 516} (1998) 449
\bibitem{n=1} S.J. Gates, Jr., {\it Nucl.Phys.} {\bf B 184} (1981) 381; 
\bibitem{siegel} W. Siegel, M. Ro\v{c}ek, M.T. Grisaru, S.J. Gates, Jr., {\it Superspace} ( Benjamin/Cummings Publishing Company, 1983)
 \bibitem{ten} S.J. Gates, Jr.,W. Siegel, {\it Nucl.Phys.} {\bf B 195} (1982) 39
\bibitem{cartan} S. Coleman, J. Wess, B. Zumino, {\it Phys.Rev.} {\bf 177} (1969) 2239; C. Callan, S. Coleman, J. Wess, B. Zumino, {\it Phys.Rev.} {\bf 177} (1969) 2247
\bibitem{forms} C.W. Misner, K.S. Thorne, and J.A. Wheeler, {\it Gravitation} (W.H. Freeman and Company, 21st ed. 1998) 
\bibitem{superforms} B. Zumino, Supersymmetry, proc. of a conf. ({\it Gauge Theories and Modern Field Theory}) at Northeastern University, Sept. 26,27, 1975, Boston
 eds.  R. Arnowitt and P. Nath, p.255; P. Nath, Supersymmetry and Gauge Supersymmetry, (same as above), p.281; D.V. Volkov and V.A. Soroka, {\it Teor.Mat.Fiz.} {\bf 20} (1974) 291; V.P. Akulov, V.A. Soroka and D.V. Volkov, {\it JETP Lett.} {\bf 22} (1975) 187
\bibitem{wess} J. Wess and J. Bagger, {\it Supersymmetry and supergravity} (Princeton University Press, 2nd ed. 1991)
\bibitem{trans} S.M. Kuzenko, {\it Int.J.Mod.Phys.} {\bf A 14} (1999) 1737, hep-th/9806147 (1998)
\end{thebibliography}
\end{document}